\documentclass[twocolumn]{aastex62}

\usepackage{xspace}
\usepackage{natbib}
\usepackage{verbatim}
\usepackage{graphicx,epstopdf}
\usepackage{longtable}
\usepackage{fancyref}
\usepackage{hyperref}
\usepackage{breakurl}
\usepackage{amsmath}

\newcommand{\target}{K2-288\xspace}
\newcommand{\primary}{K2-288A\xspace}
\newcommand{\secondary}{K2-288B\xspace}
\newcommand{\planet}{K2-288Bb\xspace}

\newcommand{\kep}{{\it Kepler}\xspace}
\newcommand{\ktwo}{{\it K2}\xspace}
\newcommand{\Spitzer}{{\it Spitzer}\xspace}

\newcommand{\msun}{$M_\odot$\xspace}
\newcommand{\rsun}{$R_\odot$\xspace}

\newcommand{\Teff}{$T_\mathrm{eff}$\xspace}

\newcommand{\Kepler}{\textit{Kepler}\xspace} 
\newcommand{\TERRA}{\texttt{TERRA}\xspace}
\newcommand{\ktwophot}{\texttt{k2phot}\xspace}

\newcommand{\vespa}{\texttt{vespa}\xspace}

\newcommand{\nsf}{NSF Graduate Research Fellow}

\received{}
\revised{}
\accepted{}

\begin{document}

\shorttitle{A Small Temperate Planet Discovered by Citizen Scientists}
\shortauthors{Feinstein et al.}

\author{Adina~D.~Feinstein}
\affiliation{Department of Physics and Astronomy, Tufts University, Medford, MA USA}
\affiliation{Department of Astronomy and Astrophysics, University of
Chicago, 5640 S. Ellis Ave, Chicago, IL 60637, USA}

\author{Joshua~E.~Schlieder}
\affiliation{Exoplanets and Stellar Astrophysics Laboratory, Code 667, NASA Goddard Space Flight Center, Greenbelt, MD USA}

\author{John~H.~Livingston}
\affiliation{Department of Astronomy, University of Tokyo, 7-3-1 Hongo, Bunkyo-ku, Tokyo, 113-0033, Japan}

\author{David~R.~Ciardi}
\affiliation{Caltech/IPAC-NASA Exoplanet Science Institute, M/S 100-22, 770 S. Wilson Ave, Pasadena, CA 91106 USA}

\author{Andrew~W.~Howard}
\affiliation{California Institute of Technology, 1200 E California Boulevard, Pasadena, CA, 91125, USA}

\author{Lauren Arnold}
\affiliation{Center for Marine and Environmental Studies, University of the Virgin Islands, Saint Thomas, United States Virgin Islands, USA}

\author{Geert Barentsen}
\affiliation{NASA Ames Research Center, Moffett Blvd, Mountain View, CA 94035, USA}
\affiliation{Bay Area Environmental Research Institute, 625 2nd St Ste. 209, Petaluma, CA 94952}

\author{Makennah Bristow}
\affiliation{Department of Physics, University of North Carolina at Asheville, Asheville, NC 28804, USA}

\author{Jessie L. Christiansen}
\affiliation{Caltech/IPAC-NASA Exoplanet Science Institute, M/S 100-22, 770 S. Wilson Ave, Pasadena, CA 91106 USA}

\author{Ian J. M. Crossfield}
\affiliation{Department of Physics, Massachusetts Institute of Technology, Cambridge, MA USA}

\author{Courtney D. Dressing}
\affiliation{University of California at Berkeley, Berkeley, CA 94720, USA}

\author{Erica J. Gonzales}
\altaffiliation{\nsf}
\affiliation{Department of Astronomy and Astrophysics, University of California, Santa Cruz, CA 95064, USA}

\author{Molly Kosiarek}
\altaffiliation{\nsf}
\affiliation{Department of Astronomy and Astrophysics, University of California, Santa Cruz, CA 95064, USA}


\author{Grant Miller}
\affiliation{Department of Physics, University of Oxford, Denys Wilkingson Building, Keble Road, Oxford, OX1 3RH, UK}

\author{Farisa Y. Morales}
\affiliation{Jet Propulsion Laboratory, California Institute of Technology, 4800 Oak Grove Drive, Pasadena, CA 91109, USA}
\affiliation{Department of Physics and Astronomy, Moorpark College, 7075 Campus Road, Moorpark, CA 93021, USA}

\author{Erik A. Petigura}
\affiliation{Division of Geological and Planetary Sciences, California Institute of Technology, Pasadena, CA USA}

\author{Beverly Thackeray}
\altaffiliation{\nsf}
\affiliation{Department of Astronomy, University of Maryland College Park, College Park, MD USA}

\author{Joanne Ault}
\author{Elisabeth Baeten}
\author{Alexander F. Jonkeren}
\author{James Langley}
\author{Houssen Moshinaly}
\author{Kirk Pearson}
\author{Christopher Tanner}
\author{Joanna Treasure}
\affiliation{Citizen Scientists, c/o Zooniverse, Department of Physics, University of Oxford, Denys Wilkingson Building, Keble Road, Oxford, OX1 3RH, UK}


\title{\planet: A Small Temperate Planet in a Low-Mass Binary System Discovered by Citizen Scientists}

\begin{abstract}

Observations from the \kep and \ktwo missions have provided the astronomical community with unprecedented amounts of data to search for transiting exoplanets and other astrophysical phenomena. Here, we present \target, a low-mass binary system (M2.0 $\pm$ 1.0; M3.0 $\pm$ 1.0) hosting a small (R\textsubscript{p} = 1.9 R\textsubscript{$\Earth$}), temperate (T\textsubscript{eq} = 226K) planet observed in \ktwo Campaign 4. The candidate was first identified by citizen scientists using Exoplanet Explorers hosted on the Zooniverse platform. Follow-up observations and detailed analyses validate the planet and indicate that it likely orbits the secondary star on a 31.39 day period. This orbit places \planet in or near the habitable zone of its low-mass host star. \planet resides in a system with a unique architecture, as it orbits at $>$0.1 AU from one component in a moderate separation binary ($a_{proj} \sim$55 AU), and further follow-up may provide insight into its formation and evolution. Additionally, its estimated size straddles the observed gap in the planet radius distribution. Planets of this size occur less frequently and may be in a transient phase of radius evolution. \target is the third transiting planet identified by the Exoplanet Explorers program and its discovery exemplifies the value of citizen science in the era of \Kepler, \ktwo, and TESS.

\end{abstract}


\section{Introduction} \label{sec:intro}

With the discovery and validation of over 300 planets spanning the ecliptic as of September 2018, the now retired \ktwo Mission has continued the exoplanet legacy of \kep by providing high-cadence continuous light curves for tens of thousands of stars for more than a dozen $\sim$80 day observing campaigns \citep[][]{howell2014, NEA}. The surge of data, with calibrated target pixel files from each campaign being publicly available approximately three months post-observing, is processed and searched by the astronomy community for planetary transits. However, due to spacecraft systematics and non-planetary astrophysical signals (e.g. eclipsing binaries, pulsations, etc.) that could be flagged as potential planets, all transiting candidates are vetted by-eye before proceeding with follow-up observations to validate and characterize the system.

Because thousands of signals are flagged as potential transits, by-eye vetting is a necessary, however tedious, task \citep[e.g.][]{crossfield2016, yu2018, crossfield2018}. Transits can also be missed and the lowest signal-to-noise events are often not examined. This presents the opportunity to source the search for transiting planets and other astrophysical variables in \ktwo data to the public, leveraging the innate human ability for pattern recognition and interest to be involved in the process of exoplanet discovery. The Planet Hunters\footnote{https://www.planethunters.org/} citizen science project \citep[]{fischer2012, schwamb2012}, hosted by the Zooniverse platform \citep[]{lintott2008}, pioneered the combination of {\it Kepler} and \ktwo time series data and crowd sourced searches for exoplanets and other time variable phenomena. Planet Hunters has been hugely successful, with more than ten refereed publications presenting discoveries of new planet candidates, planets, and variables (e.g. \cite{gies2013, schmitt2014, wang2013}); this includes surprising discoveries such as the enigmatic ``Boyajian's Star'' \citep[KIC 8462852,][]{boyajian2016} as well. 

Building on the success of Planet Hunters, the Exoplanet Explorers\footnote{https://www.zooniverse.org/projects/ianc2/exoplanet-explorers} program invites citizen scientists to discover new transiting exoplanets from \ktwo. Exoplanet Explorers presents processed \ktwo time series photometry with potential planetary transits as a collage of simple diagnostic plots and asks citizen scientists to cycle through the pre-identified candidates and select those matching the expected profile of a transiting exoplanet. Flagged candidates are then examined by the Exoplanet Explorers team and the most promising are prioritized for follow-up observations to validate the systems. When Exoplanet Explorers was launched, candidate transits were uploaded as soon as planet searches in new \ktwo campaigns had completed and citizen scientists were examining these new candidates simultaneously with our team. This process began with \ktwo Campaign 12 and Exoplanet Explorers immediately had success with its first discovery, K2-138, a system hosting five transiting sub-neptunes in an unbroken chain of near 3:2 resonances \citep[]{christiansen2018}. Another system simultaneously identified by our team and citizen scientists on Planet Hunters and Exoplanet Explorers is  K2-233, a young K dwarf hosting three small planets \citep[][]{david2018}.  Following the K2-138 discovery, we also made available candidates from \ktwo campaigns observed prior to the launch of Exoplanet Explorers. This allowed for the continued vetting of low signal-to-noise candidates and the opportunity to identify planets that may have been missed our team's vetting procedures.   

\begin{figure*}[!htb]
\begin{center}
\includegraphics[angle=0,scale=0.6,keepaspectratio=true]{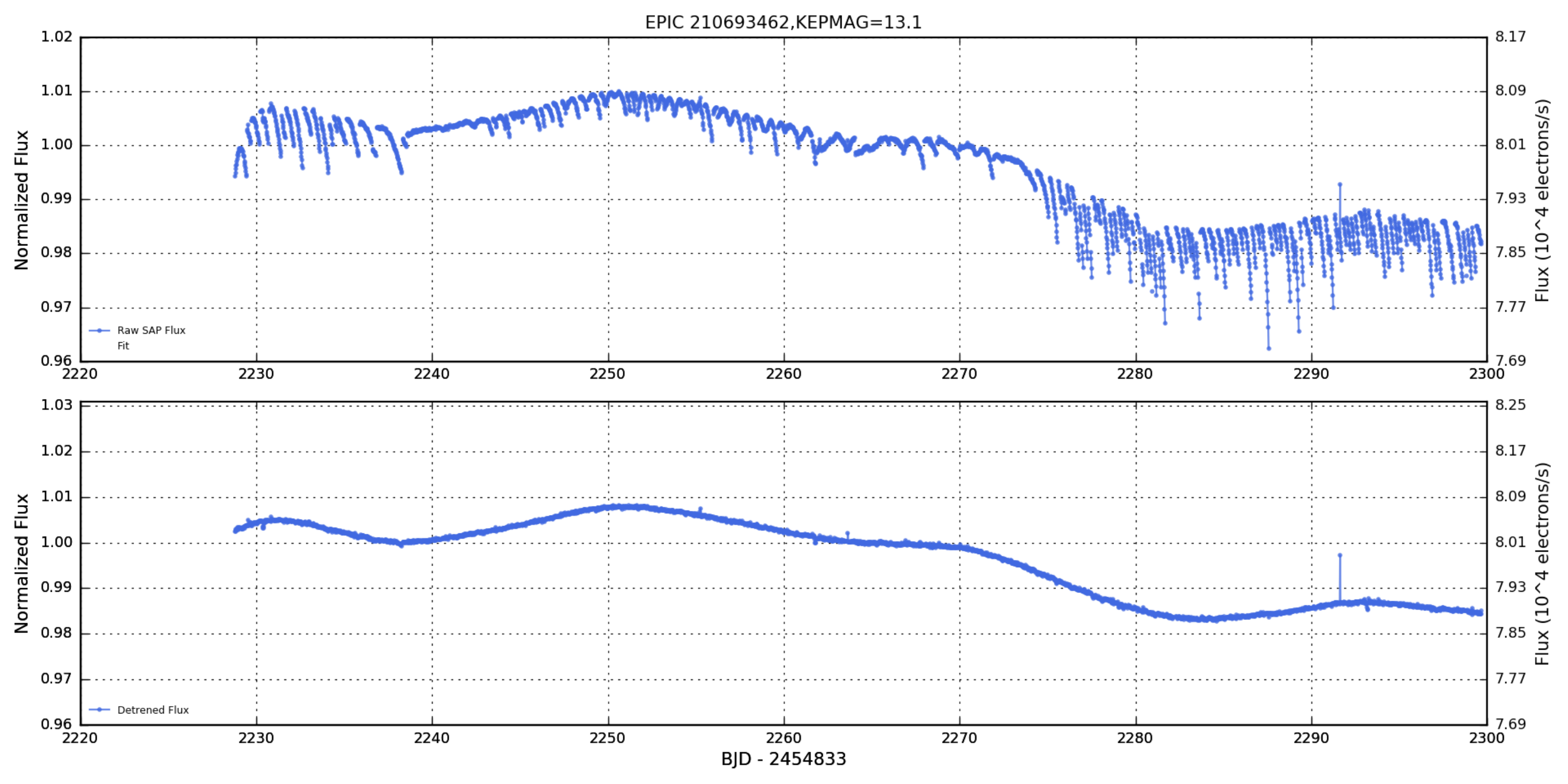}
\caption{Top: Raw \ktwo photometry of EPIC 210693462 displaying time and roll dependent spacecraft systematics. Bottom: systematics corrected, detrended light curve. \label{fig:k2phot}}
\end{center}
\end{figure*}

Here we present an example of such a system: \target from \ktwo Campaign 4, the third discovery by the citizen scientists of Exoplanet 
Explorers. \target is a small ($\sim$ 1.90$~R_\Earth$) temperate (T$\sim$226 K) planet orbiting one component of a nearby M dwarf binary. 
We layout the validation of the system in the following way: In ~\S~\ref{sec:k2}, we describe the \ktwo observations and discovery of 
the candidate by citizen scientists. We describe follow-up observations and the detection of an M-dwarf stellar secondary in ~\S~\ref{sec:floats}. In ~\S~\ref{subsec:spaceobs}, we discuss Gaia DR2 and \Spitzer follow-up observations, and we discuss transit analyses, estimated planet parameters, and system validation in ~\S~\ref{sec:discussion}. We conclude with ~\S~\ref{sec:conclusions}, which summarizes our final remarks on the system and
expresses the importance of citizen scientists for future exoplanet discoveries.


\section{K2 Observations and Candidate Identification} \label{sec:k2}

\begin{figure}[!htb]
\begin{center}
\includegraphics[angle=0,scale=0.26,keepaspectratio=true]{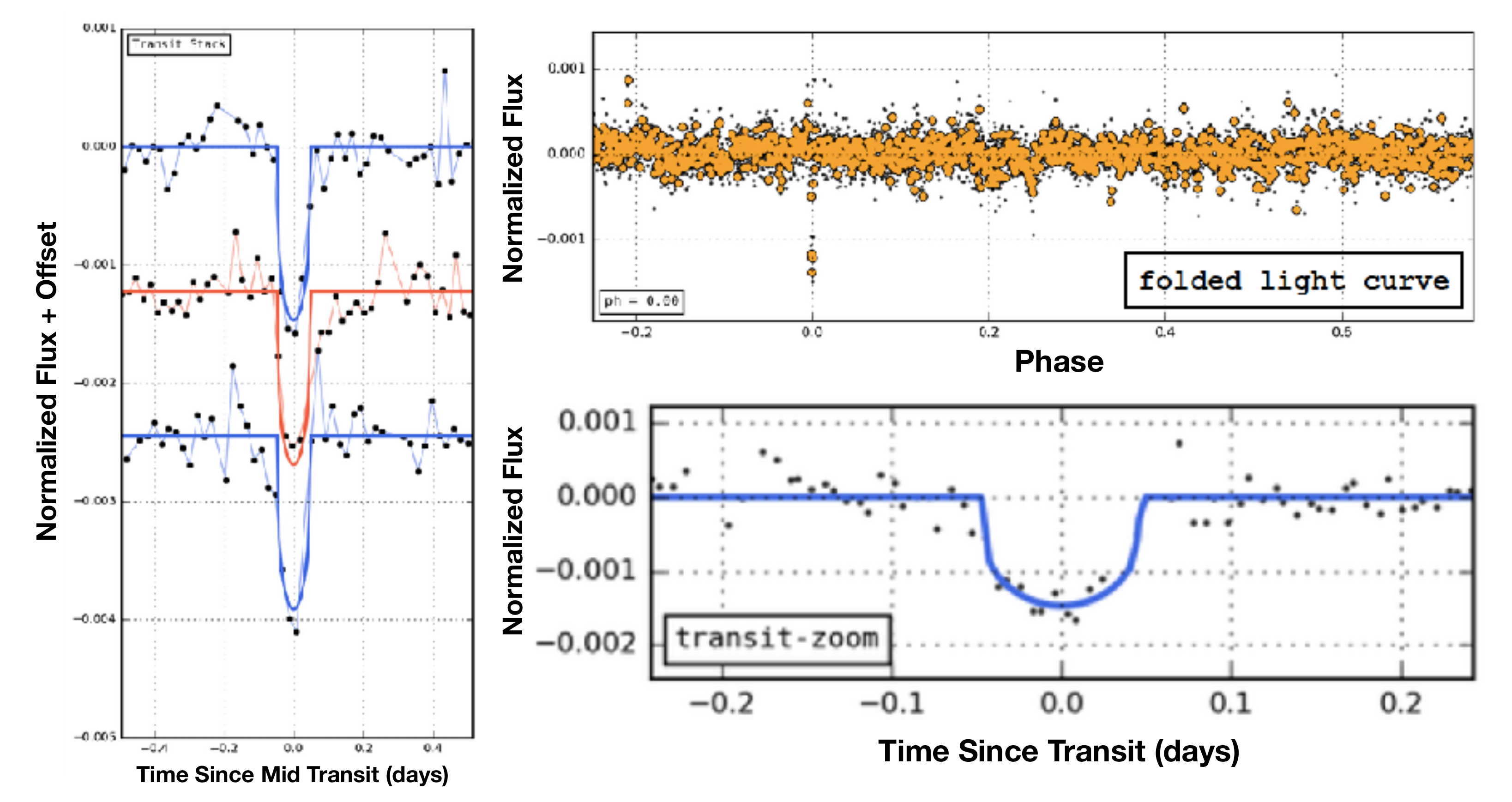}
\caption{Collage of vetting diagnostics for \target presented to citizen scientists on Exoplanet Explorers. The left image shows a stack of the individual transits with arbitrary flux offsets and alternating model fit colors for clarity. The top right image is the full \ktwo light curve folded onto the period of the transit like signal. The small black points are the \ktwo data, the orange circles are binned. The bottom right is a zoom on the transit in the period folded light curve with a preliminary planet model in blue. The detection of three transits with consistent shapes and depths and a folded transit with a planet like profile led citizen scientists to flag this event as a candidate planet. \label{fig:eeplot}}
\end{center}
\end{figure}

\target (EPIC 210693462, LP 413-32, NLTT 11596, 2MASS J03414639+1816082) was proposed as a target in \ktwo Campaign 4 (C4) by four teams
in K2 GO Cycle 1\footnote{GO4011 - PI Beichman; GO4020 - PI Stello; GO2060 PI Coughlin; GO4109 - PI Anglada}. The target was 
subsequently observed at 30-minute cadence for 75 days in C4, which ran from 2015 February 7 until 2015 April 23. Following our team's 
previous work \citep[]{crossfield2016, petigura2018, yu2018}, we used the publicly available \ktwophot software 
package\footnote{https://github.com/petigura/k2phot} \citep[]{petigura2015} to simultaneously model spacecraft systematics and stellar 
variability to detrend all C4 data. Periodic transit like signals were then identified using the publicly available \texttt{TERRA} 
algorithm\footnote{https://github.com/petigura/terra} \citep[]{petigura2013a, petigura2013b}. In this initial search of the detrended EPIC 
210693462 light curve, \TERRA did not identify any periodic signals with at least three transits. Subsequently, all C4 data was 
re-processed using an updated version of \ktwophot (see Fig.~\ref{fig:k2phot}) and searched again for transit like signals using \TERRA.
Transit candidates from these re-processed light curves were uploaded to Exoplanet Explorers. Citizen scientists participating in the 
project identified a previously unrecognized candidate transiting \target (see Fig.~\ref{fig:eeplot}). 

Citizen scientists of the Exoplanet Explorers project are presented with a portion of a \TERRA processed \ktwo light curve. The 
presentation includes a light curve folded onto the phase of the candidate transit and a stack of the individual transit events
(Fig.~\ref{fig:eeplot}). After a brief introduction, users are asked to examine the light curve diagnostic plots and select candidates that have features consistent with a transiting planet. Sixteen citizen scientists identified the candidate transiting \target as a candidate of interest. The newly identified candidate transited just three times during \ktwo C4 with a period of approximately 31 days. In the discussion forums of Exoplanet Explorers, some of the citizen scientists used preliminary stellar \citep[from the EPIC,][]{huber2016} and planet (from the \texttt{TERRA} output) parameters to estimate that the transiting candidate was approximately Earth sized and the incident stellar flux it received was comparable to the flux received by the Earth, increasing our interest in the system. 

Through Zooniverse, we contacted the citizen scientists who flagged this system as a potential transit. Many were pleasantly surprised and excited to hear that they were able to contribute to the scientific community. Additionally, they were very appreciative of our reaching out and giving them the opportunity to receive credit for their contributions and participate in this work. 50\% of those citizen scientists involved responded to our email and are included as co-authors on this publication; the rest are thanked in the acknowledgements. We aim to continue the precedent set by \cite{christiansen2018} of attributing credit to all, including citizen scientists, who are involved in planetary system identification and validation.

After the discovery by citizen scientists, we investigated the full 
\ktwophot light curve and the \TERRA outputs to understand how this 
intriguing candidate was overlooked in our catalog of planets and 
candidates from the first year of \ktwo \citep[][]{crossfield2016}. 
Our investigation revealed that the 
candidate was missed by our first analysis of the \ktwo C4 light curves  
because the version of the \ktwophot software used trimmed data from the beginning 
and end of the observing sequence. This is a common practice to mitigate 
systematics at the start and finish of a \ktwo campaign. The first transit, occurring only 2 days into the observing sequence, was trimmed from the data prior to running \TERRA and the algorithm did not flag the candidate because it only transited twice (see Fig.~\ref{fig:ktwofit}). We searched additional publicly 
available light curves of \target on the Mikulski Archive for Space 
Telescopes (MAST)\footnote{https://archive.stsci.edu/k2/} for a similar 
transiting candidate. Due to data trimming similar to that applied to our original 
\ktwophot light curve, the \texttt{k2sff} \citep[]{vanderburg2014} light curve also 
exhibited only two transits and the candidate was not published in the 
catalog of \citet[]{mayo2018}. However, three transits were 
recovered in the \texttt{EVEREST} \citep[]{luger2016, luger2017} and \texttt{k2sc} \citep[]{aigrain2016} light 
curves but the candidate and its parameters have not yet been published. 
Following these checks, we compiled known information on the system (see 
Table~\ref{tab:stellarparameters}) and began 
follow-up observations to further characterize the host star and validate 
the candidate planet.


\section{Ground Based Observations} \label{sec:floats}

\subsection{IRTF SpeX}\label{subsec:spex}

The first step in our follow-up process was observing \target\ with the near-infrared cross-dispersed spectrograph, SpeX \citep[][]{rayner2003, rayner2004} on the 3-meter NASA Infrared Telescope Facility. The observations were completed on 2017 July 31 UT (Program 2017A019, PI C. Dressing). The target was observed under favorable conditions, with an average seeing of $\sim$0.8$^{\prime\prime}$. We used SpeX in its short cross-dispersed mode (SXD) with the 0.3$\times$15$^{\prime\prime}$ slit, covering 0.7-2.55 $\mu$m at a resolution of R $\sim$ 2000. The target was observed for an integration time of 120s per frame at two locations along the slit in 3 AB nod pairs, leading to a total integration time of 720s. The slit position angle was aligned to the parallactic angle in order to minimize differential slit losses. After observing \target, we immediately observed a nearby A0 standard, HD23258, for later telluric correction. Flat and arc lamp exposures were also taken for wavelength calibration. The spectrum was reduced using the SpeXTool package \citep[][]{vacca2003,cushing2004}.

\begin{figure*}[!ht]
\begin{center}
\includegraphics[angle=0,scale=0.8,keepaspectratio=true]{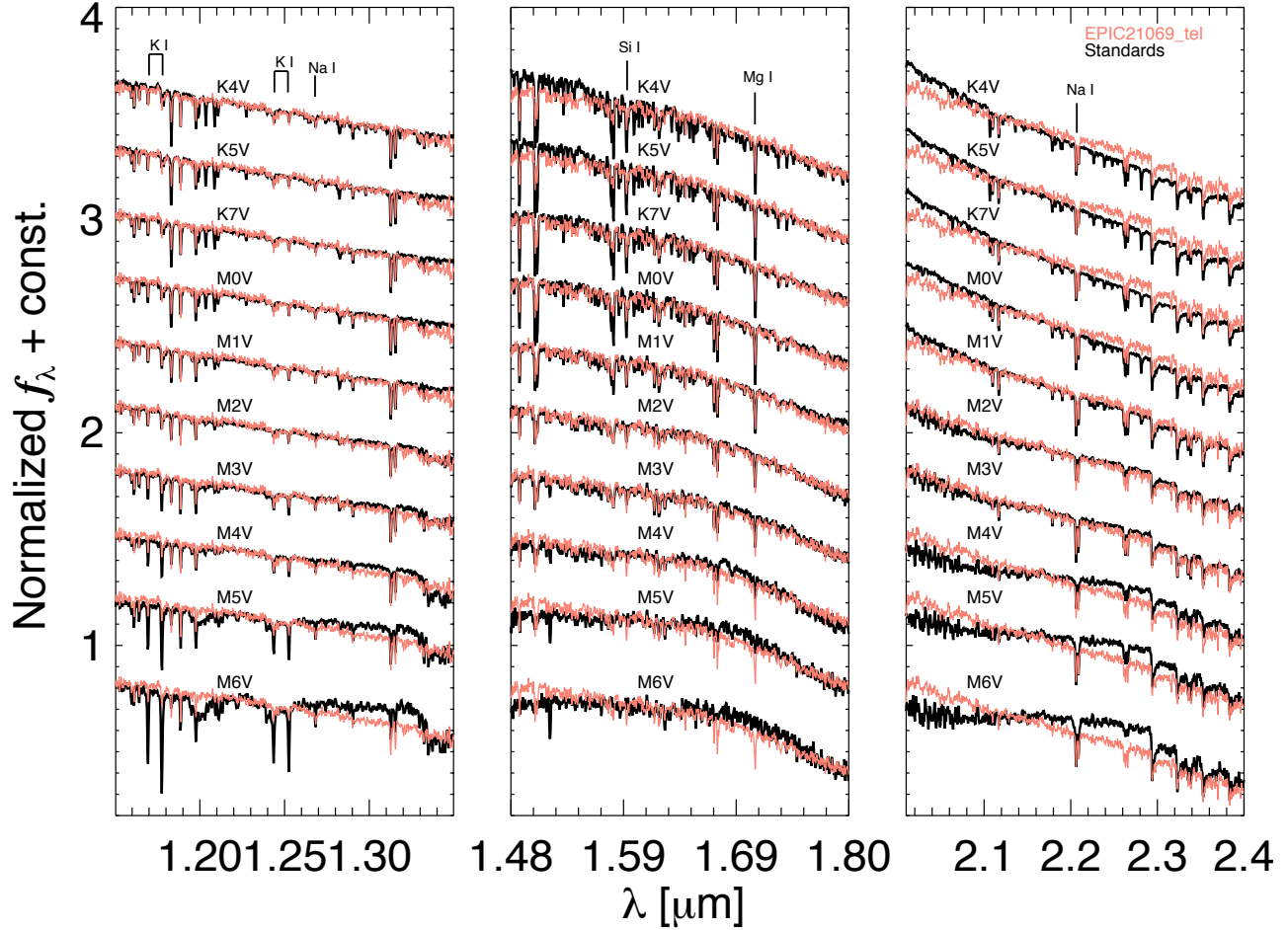}
\caption{ The $JHK$-band spectra of \target (EPIC 210693462) obtained using SpeX on the IRTF (salmon) compared to late-type dwarf standards from the IRTF spectral library (black). All spectra are normalized to the continuum in each of the plotted regions. After processing using SpeXTool, the resulting spectra is a best visual match to the M2/M3 spectral type across the three $JHK$-bands. \label{fig:spex}}
\end{center}
\end{figure*}

SpeXTool uses the obtained target spectra, A0 standard spectra, and flat and arc lamp exposures to complete the following reductions: flat fielding, bad pixel removal, wavelength calibration, sky subtraction, and flux calibration. The package yields an extracted and combined spectra. The resulting two spectra have signal-to-noise ratios (SNRs) of 106 in the \textit{J}-band ($\sim$1.25 $\mu$m), 127 in the \textit{H}-band ($\sim$1.6 $\mu$m), and 107 per resolution in the \textit{K}-band ($\sim$2.2 $\mu$m). The reduced spectra is compared to late-type standards from the IRTF Spectral Library \citep[]{rayner2009} across the \textit{JHK}-bands in Figure ~\ref{fig:spex}. Upon visual inspection, \target\ is an approximate match to the M2/M3 standard across all three bands. This is consistent with the spectral type estimated using the NIR index based H$_20_{K2}$ method of \citet{rojas2012}, M2.0 $\pm$ 0.6, and the optical index based TiO5 and CaH3 methods of \citet{lepine2013}, M3.0 $\pm$ 0.5. 

We used the SpeX spectrum to approximate the fundamental parameters of the star (metallicity, [Fe/H]; effective temperature, \Teff; radius, $R_*$; mass, $M_*$; and luminosity, $L_*$) following the prescription presented in \citet{dressing2017a}. Specifically, we estimate the stellar \Teff, $R_*$, and $L_*$ using the relations of \citet{newton2015}, the metallicity using the relations of \citet{mann2013a}, and $M_*$ by using the \citet{newton2015} \Teff in the temperature-mass relation of \citet{mann2013b}. \citet{newton2015} used a sample of late-type stars with measured radii and precise distances to develop a relationship between the equivalent widths of $H$-band Al and Mg lines and fundamental parameters. \citet{mann2013a} used a set of wide binaries with solar type primaries and M dwarf companions to calibrate a relationship between metallicity and the strength of metallicity sensitive spectroscopic indices.  \citet{mann2013b} derived an empirical effective temperature relationship using a sample of low-mass stars with measured radii and distances and temperature sensitive indices in the near infrared spectra of low-mass stars. Using the same samples, they then derived additional empirical relations for \Teff-$R_*$, \Teff-$M_*$, and \Teff-$L_*$. Our application of these empirical relationships to the SpeX spectrum of \target following the prescription of \citet{dressing2017a} results in \Teff = 3479 $\pm$ 85 K, $R_*$ = 0.47 $\pm$ 0.03 \rsun, $M_*$ = 0.38 $\pm$ 0.08 \msun, log($L/L_*$) = -1.53 $\pm$ 0.06, and [Fe/H] = -0.06 $\pm$ 0.21. The estimated stellar parameters are consistent with the M2.5 spectral type measured from the spectrum. We note that these values apply to the blended spectrum of a binary system and are not indicative of the final stellar parameters for the components in the system. We discuss the discovery and properties of the binary in ~\S ~\ref{subsec:hires}, \ref{subsec:hri}, and~\ref{sec:properties}.

\subsection{Keck HIRES}\label{subsec:hires}

We observed \target on 2017 Aug 18 UT with the HIRES spectrometer \citep[]{vogt1994} on the Keck I telescope. The star was observed following the standard California Planet Survey \citep[CPS,][]{marcy2008, howard2010} procedures with the C2 decker, $0\farcs87 \times 14\farcs0$ slit, and  no iodine cell. This set-up provides wavelength coverage from 3600 - 8000~\AA~at a resolution of $R\approx60,000$. We integrated for 374s, achieving 10,000 counts on the HIRES exposure meter for an SNR of $\sim$25 per pixel at 5500~\AA. The target was observed under favorable conditions, with seeing $\sim$ 1$^{\prime\prime}$. During these observations, we noted that the intensity distribution of the source in the HIRES guider images was elongated approximately along the SE-NW axis. We observed \target again on 2017 Aug 19 UT using the same instrument settings and integration time, but in slightly better seeing. A secondary component was partially resolved in the guider images at $\sim$1$^{\prime\prime}$ to the SE. This observation prompted adaptive optics imaging using Keck NIRC2 to fully resolve the binary (see \S~\ref{subsec:hri}). Following the identification of the secondary, we observed \target again with Keck HIRES on 2017 Sept 6 UT with an integration time of 500s in $< 1^{\prime\prime}$ seeing, achieving an SNR $\sim$20 per pixel at 5500~$\AA$. During these observations, we oriented the slit to be perpendicular to the binary axis (PA = 330$^{\circ}$) and shifted the slit position to center it on the secondary (\secondary). All HIRES spectra were reduced using standard routines developed for the California Planet Survey \citep[CPS]{howard2010}.

Visual inspection of the reduced blended and secondary spectra revealed morphologies and features consistent with low activity M dwarfs. All spectra exhibited H$\alpha$ absorption, with no discernible emission in the line cores or wings. Weak emission cores were visible in the Ca II H\&K lines, however we did not measure their strengths due to the very low SNR ($\le$3) at short wavelengths. Such weak emission is often observed even in low-activity M dwarfs. We derived stellar parameters from the spectra using the SpecMatch-Emp code \citep[]{yee2017}\footnote{https://github.com/samuelyeewl/specmatch-emp}. SpecMatch-Emp is a software tool that uses a diverse spectral library of $\sim$400 well-characterized stars to estimate the stellar parameters of an input spectrum. The library is made up of HIRES spectra taken at high SNR($>$100 per pixel). SpecMatch-Emp finds the optimum linear combination of library spectra that best matches the unknown target spectrum and interpolates the stellar \Teff, $R_*$, and [Fe/H]. SpecMatch-Emp performs particularly well on stars with  \Teff~$<$4700 K, so it is well suited to \target, a pair of M dwarfs. SpecMatch-Emp achieves an accuracy of 70 K in \Teff, 10\% in $R_*$, and 0.12 dex in [Fe/H] \citep[]{yee2017}. The library parameters are derived from model-independent techniques (i.e. interferometry or spectrophotometry) and therefore do not suffer from model-dependent offsets associated with low-mass stars \citep{newton2015, dressing2017a}. Our SpecMatch-Emp analysis of the blended spectra resulted in mean parameters of \Teff~$ = 3593 \pm 70$ K, $R_* = 0.44 \pm 0.10 R_{\odot}$, and [Fe/H] = -0.29 $\pm$ 0.09. Consistent with an M2.0 $\pm$ 0.5 spectral type following the color-temperature conversions of \citet{pecaut2013}\footnote{Throughout this work, when we refer to the \citet{pecaut2013} color-temperature conversion table, we use the updated Version 2018.03.22 table available on E. Mamajek's website - \url{http://www.pas.rochester.edu/~emamajek/EEM_dwarf_UBVIJHK_colors_Teff.txt}}. The SpecMatch-Emp analysis of the secondary spectrum resulted in $T_{eff} = 3456 \pm 70$ K, $R_* = 0.41 \pm 0.10 R_{\odot}$, and [Fe/H] = -0.21 $\pm$ 0.09. The spectroscopic temperature of the secondary is approximately 150K cooler than the blended spectrum. This is consistent with an M3.0 $\pm$ 0.5 spectral type \citep[]{pecaut2013}. The HIRES stellar parameters for the blended spectrum are also consistent with the SpeX parameters within uncertainties. Since the HIRES spectra are blended or only partially resolved, we only use the metallicities in subsequent analyses. As expected for stars in a bound system, the metallicities from the different spectra are consistent. The measured metallicities are provided in Table~\ref{tab:stellarparameters}.  


The standard CPS analyses of the HIRES spectra also provide barycentric corrected radial velocities (RV). Our two epochs of blended HIRES spectra provide a mean RV of $73.0 \pm 0.3$ km s$^{-1}$. The partially resolved secondary spectrum yields RV = $70.2 \pm 0.3$ km s$^{-1}$. These RVs are broadly consistent but differ at the 9$\sigma$ level, potentially due to orbital motion. To search for additional stellar companions at very small separations, we performed the secondary line search algorithm presented by \cite{kolbl2015} on the HIRES spectra. This analysis did not reveal any significant signals attributable to additional unseen companions in the system at $\Delta$RV $\ge$ 10 km s$^{-1}$ and $\Delta$$V$$\lesssim$ 5 mag. We report the weighted mean HIRES RV in Table~\ref{tab:stellarparameters}. 

\subsection{High-resolution Imaging}\label{subsec:hri}

After the binary companion was identified, we observed \target with high-resolution adaptive optics (AO) imaging at the Keck Observatory. This was completed in order to ensure our transit signal was due to the presence of an exoplanet and not the stellar companion. The observations were made on Keck-II with the NIRC2 instrument behind the natural guide star AO system. These observations were completed on 2017 Aug 20 UT in the standard 3-point dither pattern used with NIRC2. This observing mode was chosen to avoid the typically noisier lower left quadrant of the detector. We observed \target\ in the narrow-band $Br-\gamma$, the H-continuum, and J-continuum filters. Using a step-size of $3\arcsec$, the dither pattern was repeated three times, with each dither offset from the previous by 0.5$\arcsec$. We used integration times of 6.6, 4.0, and 2.0s, for the the narrow-band $Br-\gamma$, the H-continuum, and J-continuum respectively, with the co-add per frame for a total of 59.4, 36.0, and 26.1s. The narrow-angle mode of the camera allowed for a full field of view of $10 \arcsec$ and a pixel scale of approximately $0.009942 \arcsec$ per pixel. The Keck AO observations clearly detected a nearly equal brightness secondary  $\sim 0.8\arcsec$ to the southeast of the primary target. We also observed \target on 2017~Dec~29 UT in the broader $J$ and $K_p$ filters through poor and variable seeing ($\sim$1-2$^{\prime\prime}$). The binary was clearly resolved, but the images were of much lower quality than the 2017~Aug~20 observations and are not used in any subsequent analyses.

\begin{figure}[!ht]
\begin{center}
\includegraphics[angle=0,scale=0.7,keepaspectratio=true]{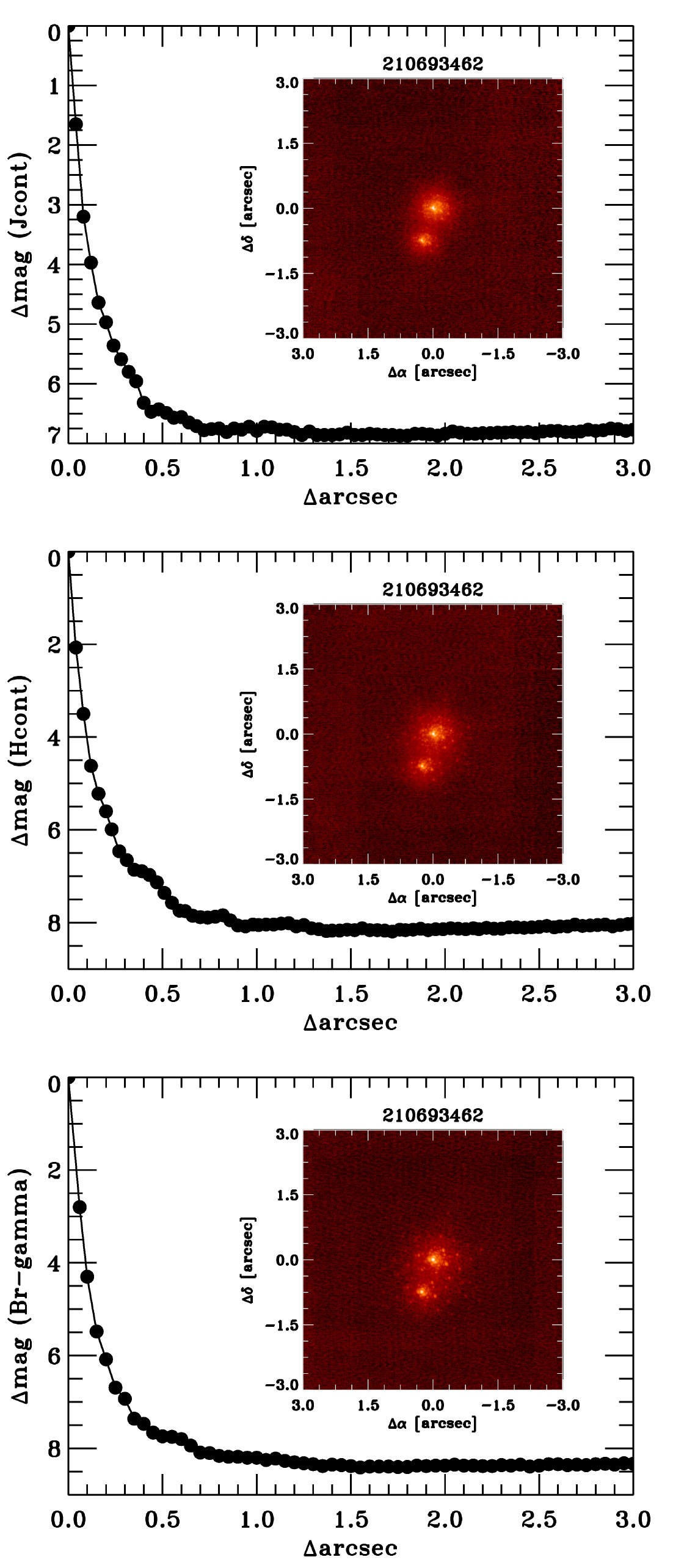}
\caption{Contrast sensitivities and Keck/NIRC2 AO images (insets) of \target in the $Jcont$, $Hcont$, and $Br-\gamma$ filters. A secondary secondary is clearly detected at $\sim$0.8" in each band. The 5$\sigma$ contrast limits for additional secondarys are plotted against angular separation in arcseconds for each filter; the black points represent one step in the FWHM resolution of the images.}
\label{fig:keckIm}
\end{center}
\end{figure}

The resulting NIRC2 AO data have a resolution of 0.049$^{\prime\prime}$ (FWHM) in the $Br$-$\gamma$ filter, 0.040$^{\prime\prime}$ (FWHM) in the $H$-$cont$, and 0.039$^{\prime\prime}$ (FWHM) in the $J$-$cont$ filter. Fake sources were injected into the final combined images with separations from the primary in multiples of the central source's FWHM in order to derive the sensitities of the data \citep{furlan2017}. The $5 \sigma$ limits on the sensitivity curves are shown in Figure \ref{fig:keckIm}. The separation of the secondary was measured from the $Br$-$\gamma$ image and determined to be $\Delta\alpha = 0.259\arcsec \pm 0.001\arcsec$ and $\Delta\delta = -0.746\arcsec \pm 0.001\arcsec$, corresponding to a position angle of PA $\approx 159.8^\circ$ east of north. The blending caused by the presence of the secondary was taken into account in the resulting analysis, to obtain the correct transit depth and planetary characteristics \citep{ciardi2015}.

The blended 2MASS $JHK$-magnitudes of the system are: $J = 10.545 \pm 0.020$ mag, $H=9.946 \pm 0.023$ mag, and $K_s = 9.724 \pm 0.018$ mag.  The primary and secondary have measured magnitude differences of $\Delta J = 0.997 \pm 0.009$ mag, $\Delta H = 0.990 \pm 0.005$ mag, and $\Delta K_s = 0.988 \pm 0.004$ mag. $Br$-$\gamma$ has a central wavelength that is sufficiently close to $Ks$ to enable the deblending of the 2MASS magnitudes into the two components.  The primary star has deblended real apparent magnitudes of $J_1 = 10.910 \pm 0.027$ mag, $H_1 = 10.313 \pm 0.021$ mag, and $Ks_1 = 10.092 \pm 0.023$ mag, corresponding to $(J-H)_1 = 0.597 \pm 0.033$ mag and $(H-K_s)_1 = 0.221 \pm 0.031$ mag. The secondary star has deblended real apparent magnitudes of $J_2 = 11.907 \pm 0.0269$ mag, $H_2 = 11.303 \pm 0.021$ mag, and $Ks_2 = 11.079 \pm 0.023$ mag, corresponding to $(J-H)_2 = 0.604 \pm 0.033$ mag and $(H-K_s)_2 = 0.223 \pm 0.031$ mag. We derived the approximate deblended \textit{Kepler} magnitudes of the two components using the $(Kepmag - Ks)\ vs.\ (J-Ks)$ color relationships described in \cite{howell2012}. The resulting deblended Kepler magnitudes are $Kep_1 = 13.46\pm0.09$ mag for the primary and $Kep_2 = 14.49\pm0.10$ mag for the secondary, with a resulting Kepler magnitude difference of $\Delta Kep = 1.03\pm0.12$ mag. These deblended magnitudes were used when fitting the light curves and deriving true transit depth.

\begin{figure*}[!ht]
\begin{center}
\includegraphics[angle=0,scale=0.5,keepaspectratio=true]{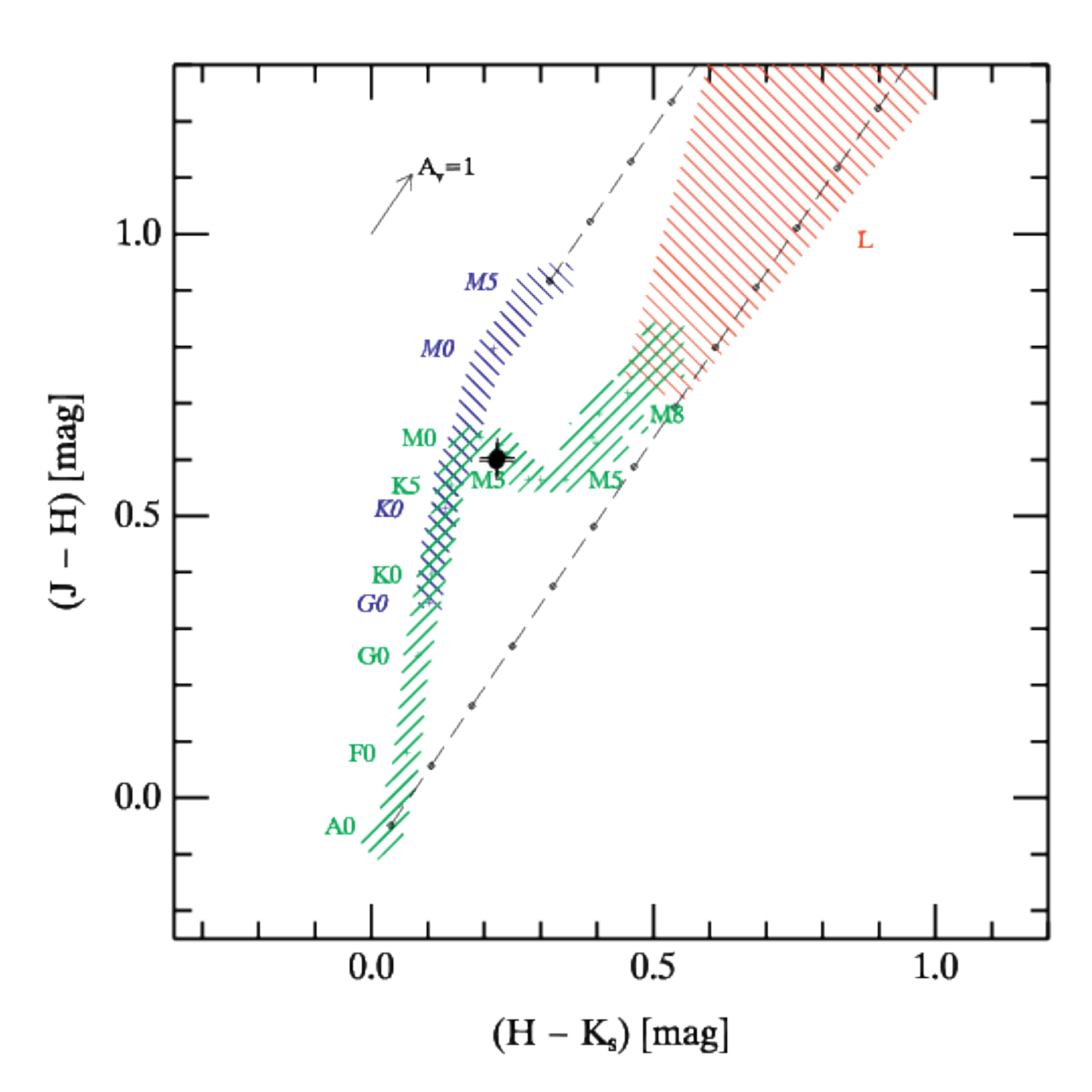}
\caption{2MASS $JHK_s$ color-color diagram of the dwarf branch locus (green; \citet{carpenter2001}, \citet{hawley2002}), the giant branch locus (blue; \citet{carpenter2001}), and the brown dwarf locus (red; \citet{kirkpatrick2000}, \citet{burgasser2002}). $A_V$, the direction of reddening due to extinction, is represented by the black dashed lines. The positions of the \target components are over plotted. The primary and secondary are both consistent with $\sim$M3V spectral types.}
\label{fig:cc}
\end{center}
\end{figure*}

Both stars have infrared colors that are consistent with approximately M3V spectral type (Figure~\ref{fig:cc}). However, this is driven by the uncertainties on the component photometry. With an approximate primary spectral type of M2,  and $\Delta$$JHK$$\approx$1 mag, the secondary is likely about one and half sub-types later than the primary \citep[]{pecaut2013}. It is unlikely that the star is a heavily reddened background star. Based on an extinction law of $R=3.1$, an early-K type star would have to be attenuated by more than 1 magnitude of extinction for it to appear as a mid M-dwarf. The line-of-sight extinction through the Galaxy is only $A_V\approx0.7$~mag at this location \citep{sf2011}, making a highly reddened background star unlikely compared to the presence of a secondary companion. Additionally, archival ground based imaging does not reveal a stationary or slow moving point source near the current location of \target, indicating that the imaged secondary at 0.8" is likely bound (See \S~\ref{subsec:archival}). Gaia DR2 also provides consistent astrometry for two stars near the location of \target (See \S~\ref{subsec:gaia}).

\subsection{Seeing Limited Archival Imaging}\label{subsec:archival}

\begin{figure*}[!ht]
\begin{center}
\includegraphics[angle=0,scale=0.58,keepaspectratio=true]{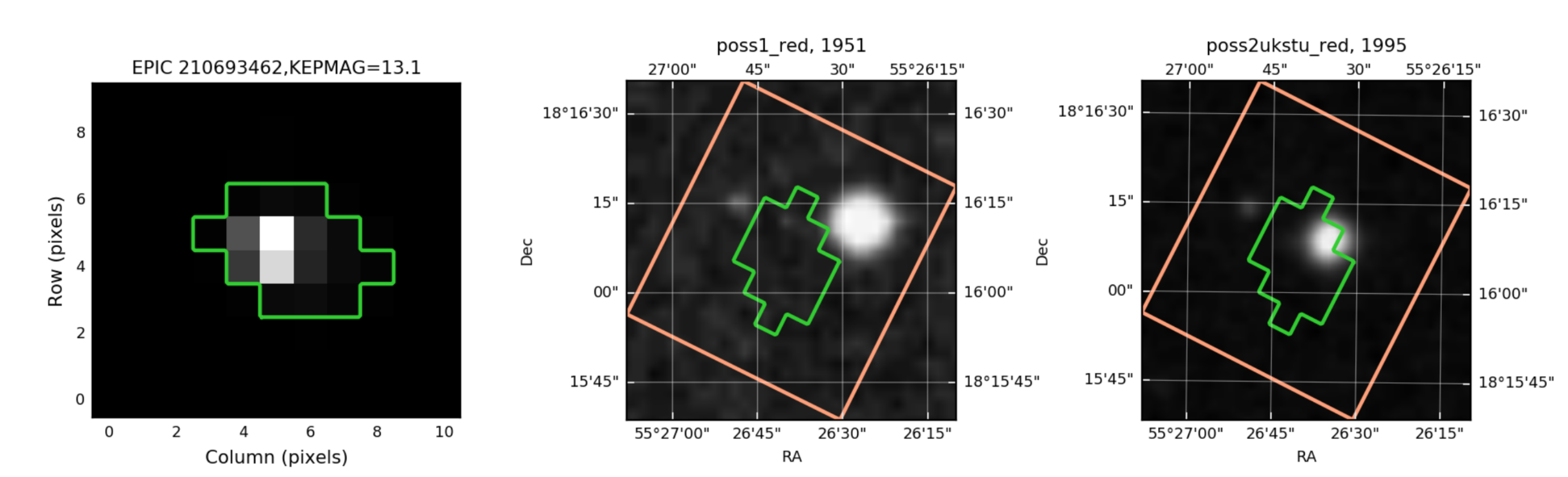}
\caption{Archival imaging data for \target. The left panel is the data from \ktwo in 2015 with the \texttt{k2phot} optimal photometric aperture overlaid in green. The center panel is a POSS I $R_{POSS}$-band image from 1951. The right panel is a POSS II $R_{POSS}$-band image from 1995. The optimal aperture is translated into the POSS image coordinates for comparison. \target has moved substantially over the nearly 65 years covered by the images. The lack of a bright star at the current position of \target in the past images indicates that there are no background sources at its current position that remain unresolved in our AO imaging observations and that the imaged companion is co-moving with the primary. We also note the faint, slow moving background source just to the NE of the optimal aperture in the POSS images. Analysis of the transit depth using different size apertures indicates that this star is not the source of the transit signal.}
\label{fig:archival}
\end{center}
\end{figure*}

\target is relatively bright and has been observed in many seeing limited surveys at multiple 
wavelengths. Currently available archival 
imaging of the system spans nearly 65 years. Over the long time baseline of the available imaging, the
large total proper motion of \target 
($\mu$ = 195.9 mas yr$^{-1}$) has carried it $\approx$12\farcs5. This allows for additional checks for very close background sources at the current location of the system and additional 
constraints on whether the resolved binary is bound or a projected background source. Figure~\ref{fig:archival} displays an image of \target from
\ktwo at its current location (left) compared to two epochs of Palomar Observatory Sky Survey (POSS) images (center and right). The green 
polygon represents the optimal photometric aperture used to extract the \ktwo light curve. In the POSS images from 1951,
there is no source at \target's current location down to the POSS I $R$ limit of 20.0 mag \citep{abell1966}. This indicates that there are no slow moving background stars that are beyond the limits of 
our AO imaging. Additionally, the lack of a bright source at the current location in archival observations
reinforces that the resolved secondary is bound and co-moving with the primary. 
    
\begin{figure*}[!ht]
\begin{center}
\includegraphics[angle=0,scale=0.55,keepaspectratio=true]{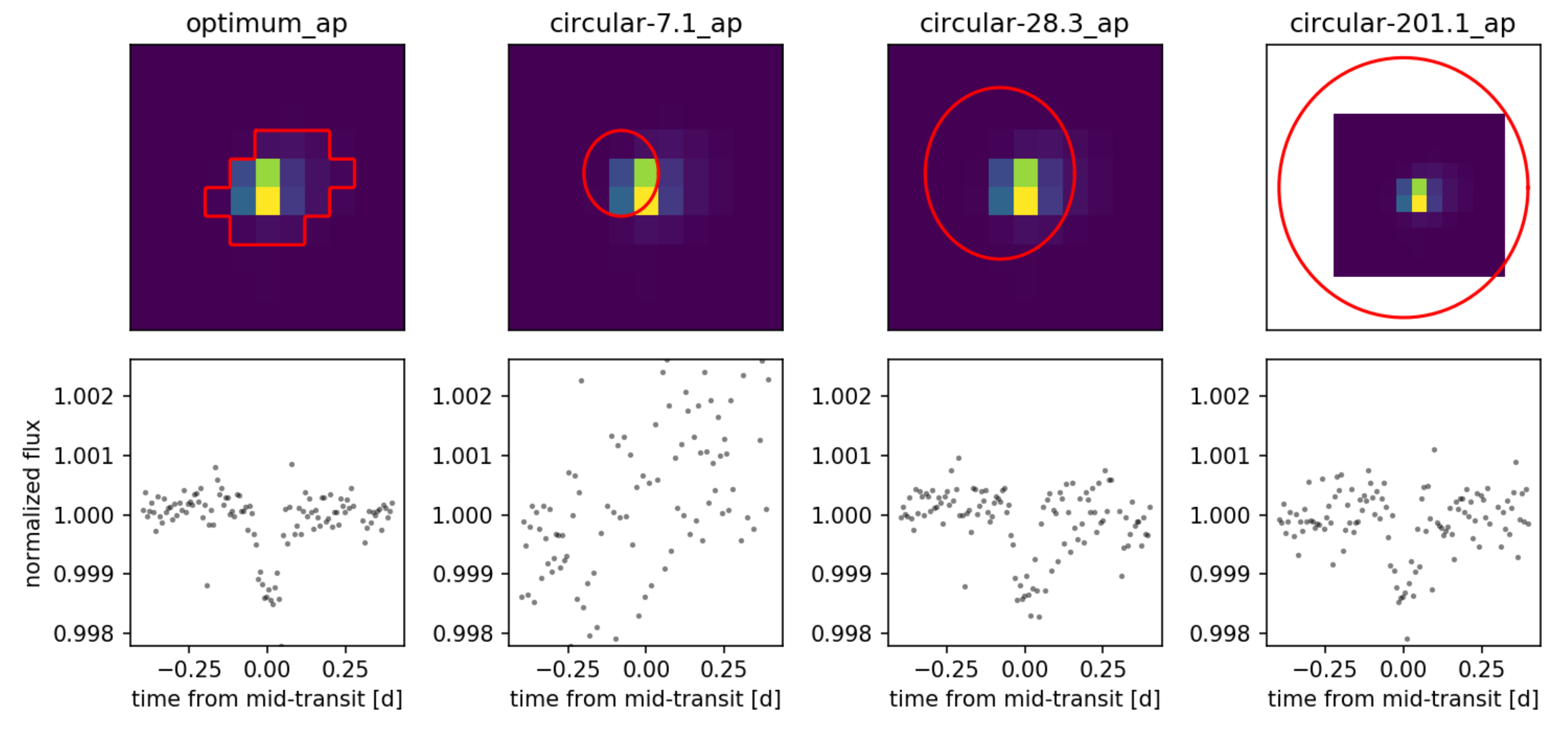}
\caption{The phase folded transit signal for \target, extracted with the optimal aperture compared and the same signal extracted using soft-edged circular apertures with radii of 1.5, 3, and 8 pixels. The proper motion of the target and the use of 2MASS coordinates to place the apertures causes the circular apertures to be slightly offset from the center. Regardless of this offset, the optimal and 3 and 8 pixel circular apertures recover transits with consistent depths and indicate that the candidate planet transits one of the components of \target, not the nearby faint background star detected in archival ground based images.}
\label{fig:aperture}
\end{center}
\end{figure*}

The archival data does reveal a faint point source $\sim$14$^{\prime\prime}$ to the NE of \target's current location. This star, 2MASS J03414730+1816135, is relatively slow moving ($\mu$ = 32.6 mas yr$^{-1}$) background source at a distance of 390 pc \citep{gaia2018}. It is $\sim$5 magnitudes fainter than \target in the \kep band and falls just outside of the optimal aperture used to produce the \ktwo light curve. Due to its proximity to the optimal aperture, this background star warrants further examination as the potential source of the transit signal. We used additional light curves generated during the \ktwophot reduction where the flux was extracted using different size apertures to investigate possible contributions from this faint star.  In Figure~\ref{fig:aperture} we show the phase folded transit signal from the light curve extracted with the optimal aperture compared to the same signal extracted using soft-edged circular apertures with radii of 1.5, 3, and 8 pixels. Due to the proper motion of the target and the use of 2MASS coordinates to place the circular apertures, the centers are offset from \target's current position by $\sim$3$^{\prime\prime}$.  None the less, the 3 and 8 pixel radius apertures yield phase folded transits with the same approximate depth as the optimal aperture while including more light from the nearby faint background star. The 1.5 pixel circular aperture suffers from a substantial increase in noise because it does not include the brightest pixels of \target. These analyses indicate that the faint slow moving background star is likely not the source of the observed transits and the candidate orbits one of the components of resolved binary. This is further reinforced by our detection of a partial transit in \Spitzer observations that use an aperture that is much smaller and free of contamination from the faint nearby star (see~\S~\ref{sec:spitzer}).

\section{Space Based Observations}\label{subsec:spaceobs}

\subsection{Gaia DR2}\label{subsec:gaia}


Astrometry \citep{Lindegren2018} and photometry 
\citep{Riello2018, Evans2018} of \target obtained by 
Gaia
over the first twenty-two months of mission operations 
were made available in the second data release from the
mission \citep[DR2][]{gaia2018}. The DR2 catalog lists 
two sources within 3\farcs7 of the 2MASS coordinates of
\target (Gaia DR2 44838019758175488 and 
44838019756570112). The separation, position angle, and
$\Delta G$ of these sources are consistent with the  
results of our Keck AO imaging and the estimated 
magnitude difference in the Kepler band. Thus, both 
components of \target were resolved by Gaia. However, 
the proximity of the sources led to relatively poor 
fits in the 5-parameter astrometric solution for each 
star. Here we refer to the goodness of fit statistic of
the astrometric solution in the along scan direction, 
{\it astrometric\_gof\_al} in the Gaia DR2 catalog.  
Good solutions typically have $astrometric\_gof\_al < 
3$, where \target A and B have values of 24.1 and 31.8,
respectively. This also leads to significant excess 
noise in the fit for each star (Gaia DR2 parameter {\it
astrometric\_excess\_noise}), 0.41 mas for the primary 
and 0.77 mas for the secondary. The utility of Gaia DR2 data in identifying binaries has been demonstrated via comparison to a large sample of AO resolved multiple systems from the \Kepler planet candidate host sample \citep{ziegler2018a}. Similarly significant excess noise in Gaia astrometric parameter fits has also been observed in this sample \citep{rizzuto2018}.

The astrometric statistics of \target may be 
improved in later Gaia data releases as more data is
obtained for each star. The excess errors are
manifested as discrepancies between the 
astrometric measurements of the components. For 
example, the parallax of the secondary differs from 
that of the primary by 0.9 mas, a $> 4 \sigma$ 
difference (when considering the secondary parallax 
uncertainty). This discrepancy is likely too large to be attributed to binary orbital 
motion over the time baseline of the Gaia observations. 

Despite this discrepancy, the distances to the components of the binary are comparable, $70.0\pm0.4$ pc and $65.7 \pm 0.9$ pc for the primary and secondary, respectively\footnote{The probabilistic distances of the components available in \cite{bailer2018} are consistent with these inverted parallax values within 0.1 pc}. Given supporting evidence that these stars form a moderate separation, bound system -- consistent RVs (see \S~\ref{subsec:hires}), consistent proper motions (see \S~\ref{subsec:archival}) -- we adopt the weighted mean and error of the primary and secondary distances as the distance to the system, $69.3 \pm 0.4$ pc, and include it in Table~\ref{tab:stellarparameters}. At this adopted distance, we find the projected separation of the secondary is $54.8 \pm 0.4$~AU. We also use 
this distance to infer the individual stellar 
parameters of the components in 
\S~\ref{sec:properties}. Gaia DR2 also provides a radial velocity for the primary, RV = $72.15 \pm 1.72$ km s$^{-1}$. This is consistent with the HIRES measured system RV and provides further evidence that there are not additional unseen stellar companions in the system.

\subsection{Spitzer Space Telescope Observations}\label{sec:spitzer}

EPIC\,210693462 was observed by \Spitzer from UT 2017-12-11 15:35:58 to 2017-12-11 20:31:28. The observations were conducted with the Infra-Red Arracy Camera (IRAC; \citealt{2004ApJS..154...10F}) at 4.5 \micron\, with an exposure time of 2 seconds. Because of the small separation of the binary components ($\sim$0.8\arcsec) and the pixel scale of IRAC (1.2\arcsec), the binary was unresolved in the \Spitzer images. Photometry of the blended binary PSF was obtained using circular apertures and the background was estimated and subtracted following a procedure similar to \citet{2016ApJ...822...39B}. The aperture was then chosen by selecting the light curve with minimal white and red noise statistics, as computed by the standard deviation and the red noise factor $\beta$ (\citealt{2006MNRAS.373..231P}, \citealt{2008ApJ...683.1076W}, Livingston et al. in review). Following this procedure, an aperture radius of 2.3 pixels was found to yield the lowest noise, which is consistent with the optimal apertures found in previous studies \citep[e.g.][]{2016ApJ...822...39B, 2012ApJ...754...22K}. We then binned the light curve and pixel data to 60 seconds, as this has been shown to yield an improved systematics correction without affecting the information content of the light curve \citep[e.g.][]{benneke2017}.

\section{System Properties \& Validation} \label{sec:discussion}

\subsection{Individual Component Properties}\label{sec:properties}

Due to the close separation of the components of 
\target, it is crucial to estimate their individual 
properties to further evaluate the 
characteristics of the planet candidate.
The spectra obtained using HIRES and SpeX are blended 
and the stellar 
parameters estimated in \S's~\ref{subsec:spex} 
and~\ref{subsec:hires} are not indicative of the 
properties of each star, except the metallicities, which should be, and are measured to be, consistent. However, our resolved NIR 
photometry from 
Keck AO imaging and the Gaia distance to the system 
provide a basis for reliably estimating the individual 
component properties. 

We base our approach on that of \cite{dressing2018sub},
which hinges on using stellar absolute $K_s$ magnitudes
($M_K$), photometric colors, and calibrated relations 
to estimate the masses, 
radii, and effective temperatures of 
low-mass stars. \cite{dressing2018sub} showed that this
approach provides fundamental parameter estimates 
consistent with those 
calculated using spectroscopic index and equivalent 
width based relations with comparable levels of 
precision. We used the 
adopted system distance of $69.3 \pm 0.4$~pc and the 
resolved $K_s$-band magnitudes of the components to 
calculate their $M_k$'s. We find $M_{K_p} = 5.888 \pm 
0.036$ mag for the primary and $M_{K_s} = 6.875 \pm 0.036$ 
mag for the secondary. Throughout the discussion, we use the subscript $p$ to denote the primary and $s$ to denote the secondary. 

To estimate the masses of the stars, we used the 
$M_K$~-~mass relation presented in \cite{benedict2016}.
We estimated stellar mass uncertainties by assuming
the errors on our absolute magnitudes 
and the coefficients in Benedict et al.~polynomial 
relation follow Gaussian distributions and calculated 
the mass 10$^4$ times using Monte Carlo (MC) methods. 
The median and standard deviation of the resulting 
distribution were adopted as the mass and associated 
statistical uncertainty.
We then added this uncertainty in quadrature to the 
intrinsic
scatter in the \cite{benedict2016} relation (0.02 
\msun). This procedure
resulted in mass estimates of $M_p = 0.52\pm 0.02 M_\odot $and$ M_s = 0.33 \pm 0.02 M_\odot$.

Our radii estimates use the $M_K$~-~radius~-~$\lbrack 
$Fe/H$ \rbrack$ relation from \cite{mann2015, 
mann2016}. In these calculations we used the HIRES measured metallicities attributed to the primary and secondary provided in Table~\ref{tab:stellarparameters}. Our approach to 
radius uncertainty estimates is similar to that used in the mass calculation. We use MC methods assuming Gaussian distributed errors on 
$M_K$ and $\lbrack $Fe/H$ \rbrack$ then add the resulting radius uncertainties in 
quadrature to the scatter in 
the \cite{mann2015, mann2016} polynomial fit (0.027 \rsun). This 
results in radii estimates of $R_p = 
0.45 \pm 0.03 R_\odot$ and $R_s = 0.32 \pm 0.03 R_\odot$.

Our effective temperature estimates use the 
$V-J$~-~$\mathrm{T_{eff}}$~-~$\lbrack $Fe/H$ \rbrack$ 
relation from \cite{mann2015, 
mann2016}. Here we also used the HIRES measured metallicities from Table~\ref{tab:stellarparameters}. The calculation also requires an estimate of the 
$V-J$ color of the stars, which we interpolate from the
\cite{pecaut2013} main 
sequence color-temperature table\footnote{We used the 
updated table
from 2018.03.22 availble on E. Mamajek's website: 
\url{http://
www.pas.rochester.edu/~emamajek/EEM_dwarf_UBVIJHK_color
s_Teff.txt}}. The $V-J$ color and uncertainty is 
estimated using MC 
methods during the interpolation. We estimate $(V-J)_p 
= 3.304 \pm 0.022$ mag for the primary and $(V-J)_s = 3.962
\pm 0.024$ mag 
for the secondary. We then used the \cite{mann2015, 
mann2016} relation to estimate the stellar temperatures
following the same 
approach to uncertainty estimation previously described
for the mass and radius estimates. We estimate the primary and secondary effective temperatures to be
$\mathrm{T_{eff, p} = 3584 \pm 205}$ 
K and $\mathrm{T_{eff, s} = 3341 \pm 
276}$ K, respectively.  These effective 
temperatures are consistent with spectral types of M2 
$\pm$ 1 and M3 $\pm$ 1 using the relations of 
\cite{pecaut2013}. Additionally, the temperature 
estimated using the resolved primary photometry is 
consistent with the temperatures estimated from the 
blended SpeX and HIRES spectra (see 
\S's~\ref{subsec:spex} and~\ref{subsec:hires}). This 
result is consistent with the $\sim$1 magnitude 
difference between the components inferred from Keck AO
imaging which reveals that the primary contributes 
substantially more flux than the secondary and dominates the blended spectra. We also use
our calculated $M_K$ 
mags and the \cite{pecaut2013} extended table to 
interpolate luminosities for \target A and B. These 
values, along with all of the other stellar parameters,
are included in Table~\ref{tab:stellarparameters}. The 
parameters estimated in this section are used in 
subsequent transit modeling analyses.

\subsection{Transit Analyses}\label{sec:transit}

\subsubsection{K2 and Spitzer Transit Modeling} \label{subsec:modeling}

\begin{figure}[!ht]
\includegraphics[width=0.45\textwidth]{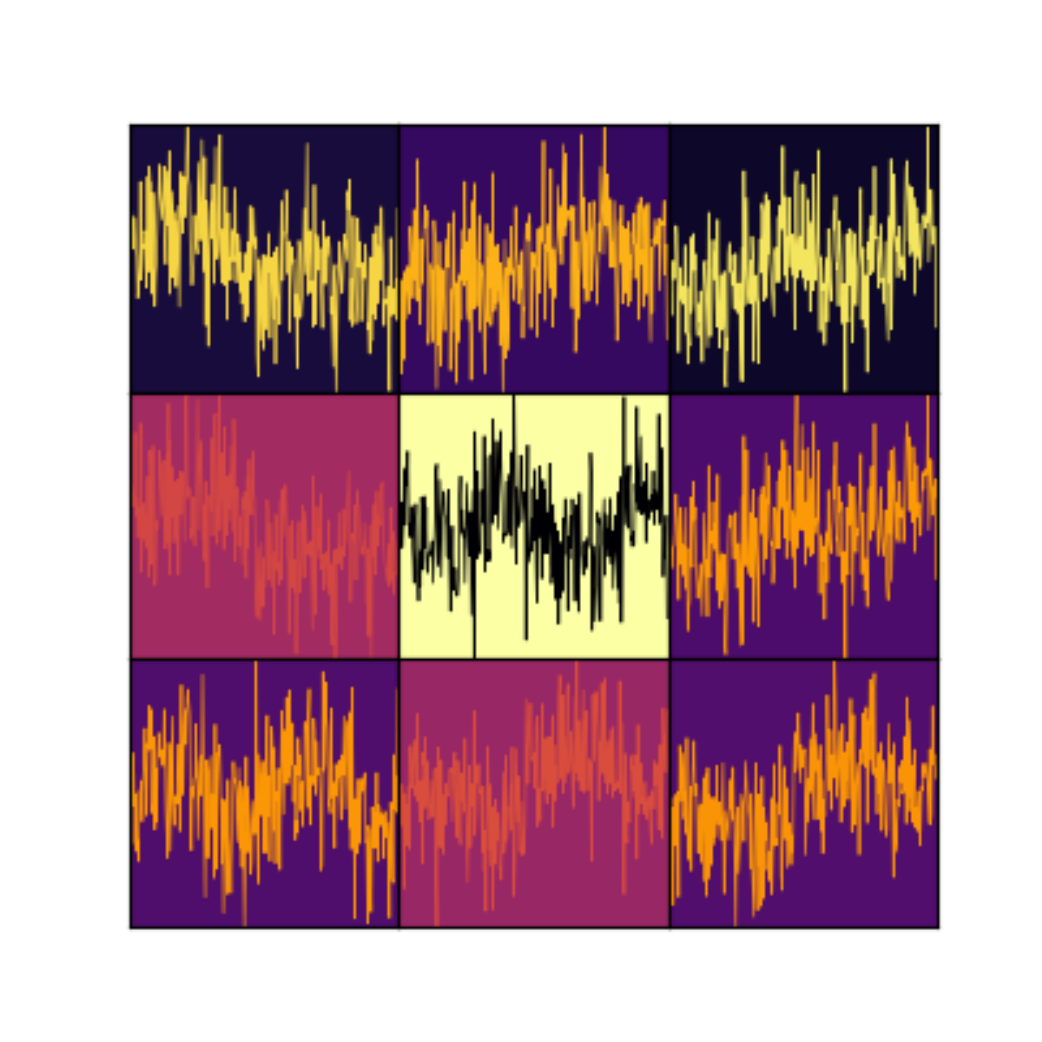}
\caption{A qualitative visualization of the individual pixel light curves used to model the Spitzer systematics via PLD, arranged in a 3$\times$3 grid and colored according to their position and flux on the IRAC detector. Each pixel light curve is normalized such that the sum of all 9 pixels at each timestep is unity (see \S\ref{subsec:modeling}); the normalization removes astrophysical variability, enabling correlated instrumental noise to be visible on multiple timescales. \label{fig:pld}}
\end{figure}

\begin{figure*}[!ht]
\begin{center}
\includegraphics[width=0.95\textwidth,trim={0.25cm 0 0 0}]{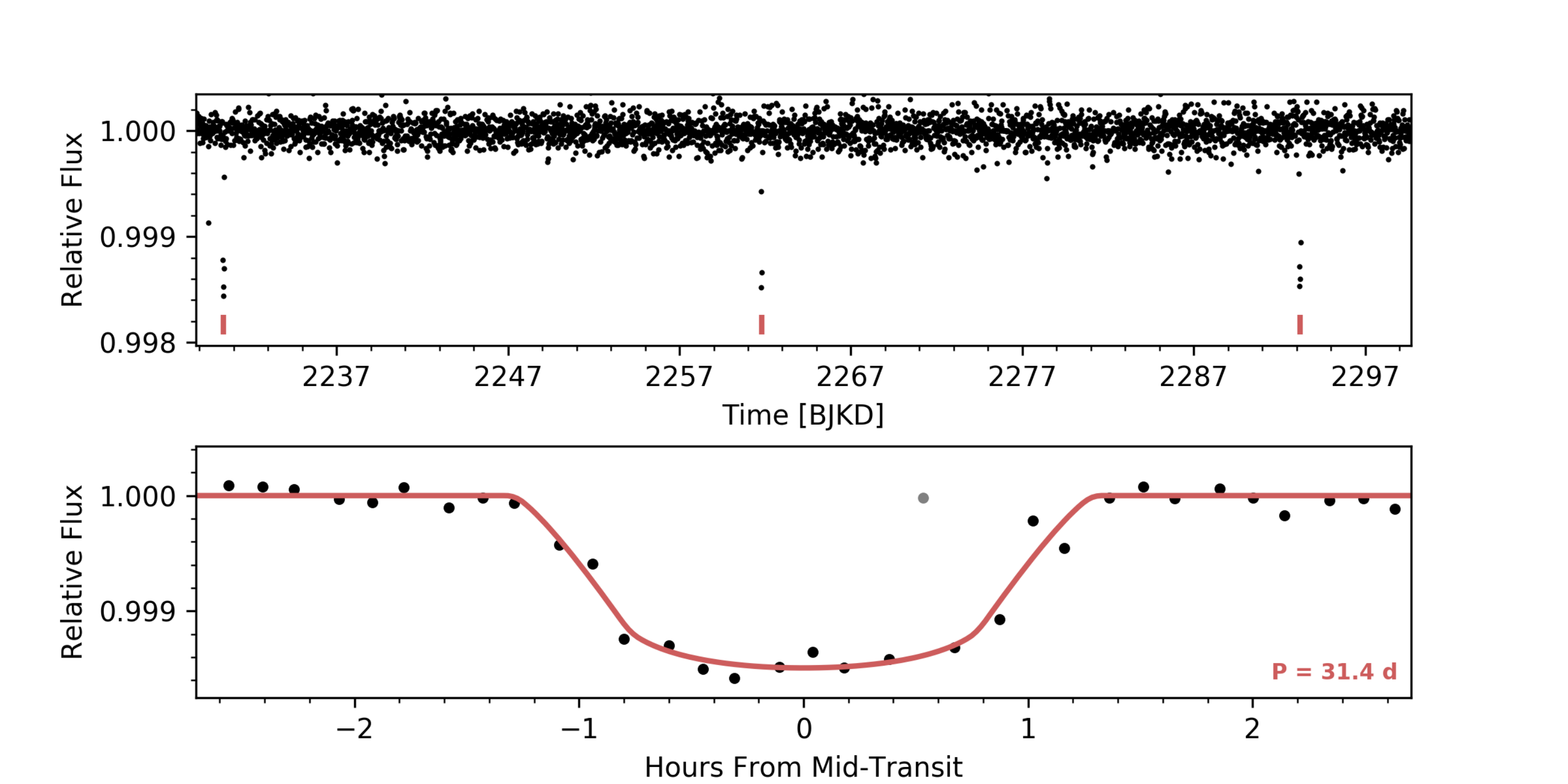}
\caption{Top: \ktwo light curve produced by EVEREST 2.0 with individual transits indicated in red. Bottom: the same light curve folded on the orbital period with best-fit transit model in red. The data point in gray was identified as an outlier and ignored in the fit. \label{fig:ktwofit}}
\end{center}
\end{figure*}

To model the \ktwo transit, we adopted a Gaussian likelihood function and the analytic transit model of \citet{mandel2002} as implemented in the Python package {\tt batman} \citep{2015PASP..127.1161K}. We used the Python package {\tt emcee} \citep{emcee} for Markov Chain Monte Carlo (MCMC) exploration of the posterior probability surface. The free parameters of the transit model are: the planet-to-star radius ratio $R_p/R_{\star}$, the scaled semi-major axis $a/R_{\star}$, mid transit time $T_0$, period $P$, impact parameter $b$, and the quadratic limb darkening coefficients $q_1$ and $q_2$ under the transformation from $u$-space of \citet{2013MNRAS.435.2152K}. The transit signal was originally identified in the \ktwophot light curve. However, for this analysis, we fit the transit model to the EVEREST 2.0 \citep{luger2017} light curve due to the lower level of residual systematics; the EVEREST 2.0 light curve and best-fit transit model are shown in Figure~\ref{fig:ktwofit}. We model the \Spitzer systematics using the pixel-level decorrelation (PLD) method proposed by \citet{2015ApJ...805..132D}, which uses a linear combination of the normalized pixel light curves to model the systematic noise caused by motion of the PSF on the detector (see Figure~\ref{fig:pld}). To allow error propagation we simultaneously model the transit and systematics using the parametrization
\begin{align}
\Delta S^t = \frac{\sum_{i=1}^9 c_i P_i^t}{\sum_{i=1}^9 P_i^t} + M_{tr}(\boldsymbol{\theta},t) + \varepsilon(\sigma),
\end{align}
where $\Delta S^t$ is the measured change in signal at time $t$, $M_{tr}$ is the transit model (with parameters $\boldsymbol{\theta}$), the $c_i$ are the PLD coefficients, $P_i^t$ is the $i^\mathrm{th}$ pixel value at time $t$, and $\varepsilon(\sigma)$ are zero-mean Gaussian errors with width $\sigma$; we fit for the logarithm of these parameters, denoted as log($\sigma$) in Table~\ref{tab:transit}. We imposed Gaussian priors on the limb darkening coefficients for both the \Kepler and IRAC2 bandpasses, with mean and standard deviation determined by propagating the uncertainties in host star properties ($T_\mathrm{eff}$, log\,$g$, and [Fe/H]) via MC sampling an interpolated grid of the theoretical limb darkening coefficients tabulated by \citet{2012yCat..35460014C}. We performed an initial fit using nonlinear least squares via the Python package {\tt lmfit} \citep{newville_2014_11813}, and then initialized 100 ``walkers'' in a Gaussian ball around the best-fit solution. We then ran an MCMC for 5000 steps and visually inspected the chains and posteriors to ensure they were smooth and unimodal, and discarded the first 3000 steps as ``burn-in.'' To ensure that we had collected enough effectively independent samples, we computed the autocorrelation time\footnote{\url{https://github.com/dfm/acor}} of each parameter. 

\begin{figure*}[!ht]
\begin{center}
\includegraphics[width=0.9\textwidth]{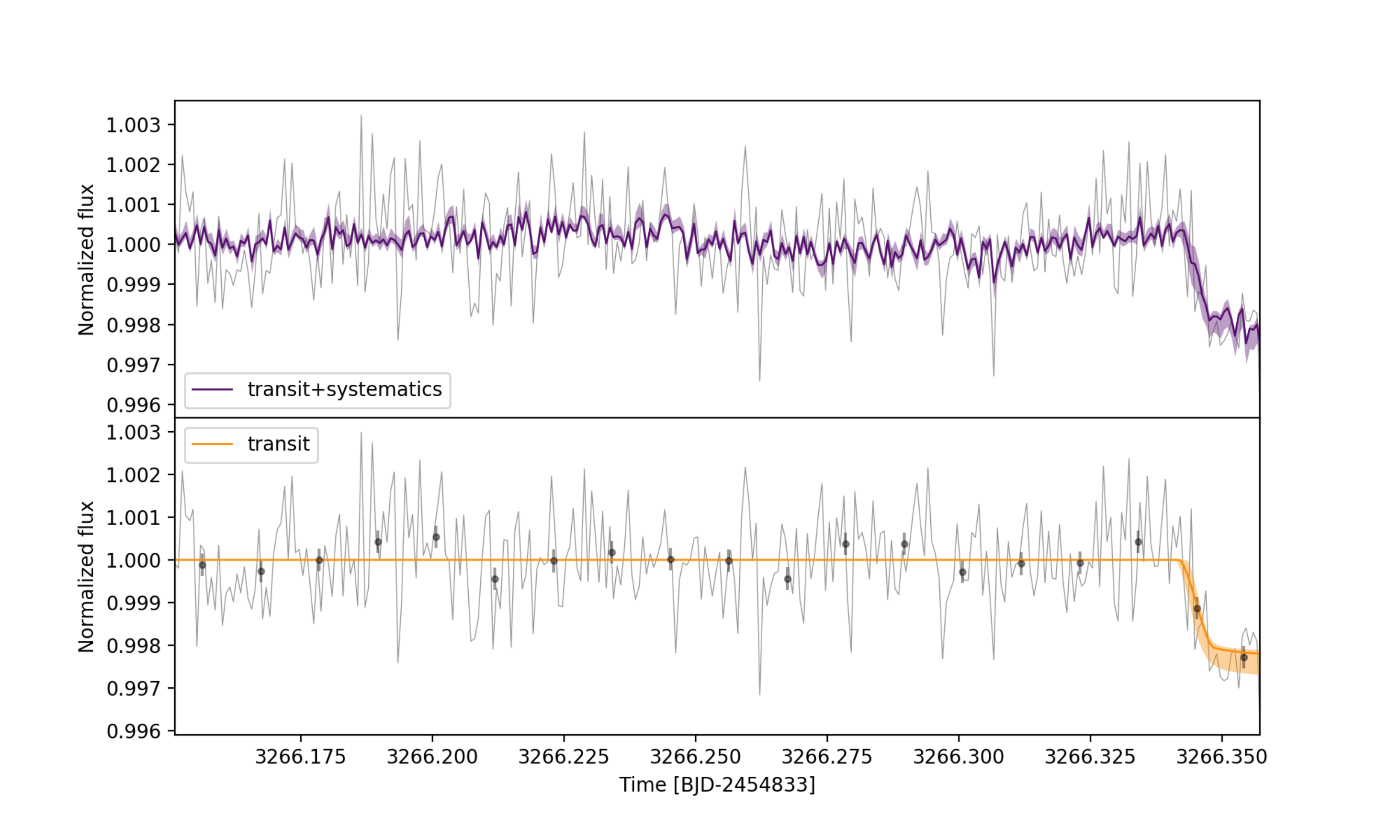}
\caption{Top: Raw \Spitzer photometry and best-fit model of the transit and systematics, with shaded 68\% credible region. Bottom: Corrected light curve and best-fit transit model, with shaded 68\% credible region. A partial transit of the planet was caught at the end of the observing sequence. \label{fig:spitzer}}
\end{center}
\end{figure*}

We plot the \Spitzer data and resulting transit fit in Figure~\ref{fig:spitzer}.  A significant partial transit, including ingress, is detected at the end of the observing sequence. The time of this transit is shifted from the transit time predicted using \ktwo data by $\sim3\sigma$. This is consistent with previous \Spitzer transit observations obtained months to years after the \ktwo observations \citep{2016ApJ...822...39B, benneke2017}. The \Spitzer observations of \target were obtained $\sim$2.5 years after \ktwo C4 and the relative  imprecision of the \ktwo transit ephemeris (a result of detecting only 3 transits) results in a significant linear drift over this time baseline (see~\citet{2016ApJ...822...39B}). We report the median and 68\% credible interval of each parameter's marginalized posterior distribution in Table~\ref{tab:transit}.

\subsubsection{Simultaneous \ktwo and \Spitzer analysis}

\begin{figure}[!ht]
\begin{center}
\includegraphics[width=0.5\textwidth,trim={0.5cm 0 0 0}]{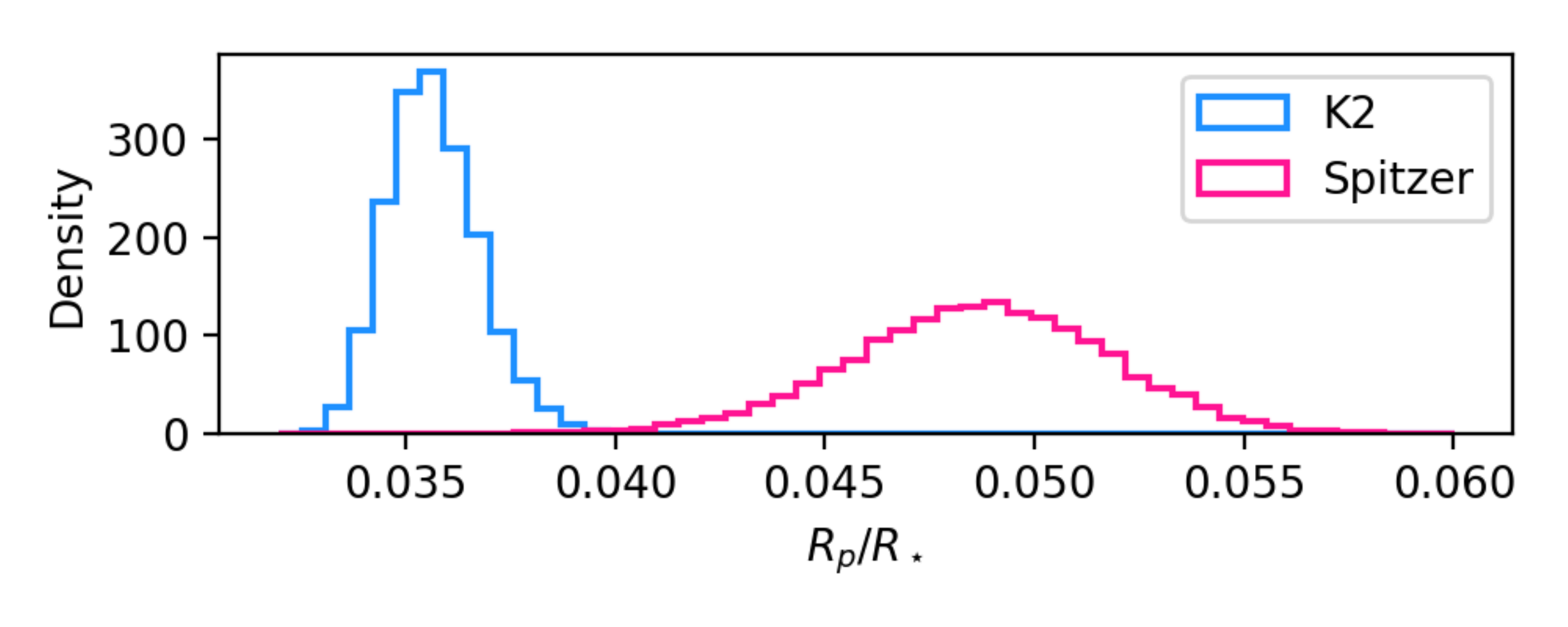}
\caption{The marginalized posterior distributions of $R_p/R_{\star}$ in the \ktwo and \Spitzer IRAC2 4.5 \micron\, bandpasses. $R_p/R_{\star}$ is significantly larger in the longer wavelength \Spitzer band, indicating the component of the binary that hosts the planet is subject to lower levels of dilution at longer wavelengths \label{fig:radii}}
\end{center}
\end{figure}

We simultaneously fit the \ktwo and \Spitzer light curves to ensure a robust recovery of the transit signal in the \Spitzer data, as well as to enable the higher cadence of the \Spitzer data to yield improved parameter estimates from the \ktwo data (Livingston et al. in review). This is achieved by sharing strictly geometric transit parameters ($a/R_{\star}$ and $b$), which are bandpass-independent, while using using separate parameters for limb-darkening, systematics, etc., which are bandpass-dependent. $R_p/R_{\star}$ is fit for both the \Kepler and \Spitzer 4.5 \micron\, bandpasses separately to allow for a dependence of transit depth on wavelength. Any such chromaticity is of particular interest in this case because it contains information about the levels of dilution present at each band, which in turn can determine which component of the binary is the true host of the transit signal. The \ktwo light curve contains only three transits, and thus only sparsely samples ingress and egress due to the 30 minute observing cadence. The impact parameter $b$ is thus poorly constrained in the fit to the \ktwo data alone, but the addition of the higher cadence \Spitzer transit yields an improved constraint, which in turn yields a more precise measurement of $R_p/R_{\star}$ in the \Kepler bandpass. A grazing transit geometry is strongly ruled out, and a significantly larger value of $R_p/R_{\star}$ is detected in the \Spitzer bandpass ($\sim$3.7$\sigma$), indicating that the component of the binary that hosts the planet candidate is subject to lower levels of dilution at longer wavelengths (see Figure~\ref{fig:radii}).

\subsection{Planet Properties and Validation}\label{sec:properties}

We derive the planet parameters for our system assuming two configurations: the planet orbits the primary M2V and the planet orbits the secondary M3V. We complete this analysis using parameters from both our \ktwo and \Spitzer transit fits. Results are presented in Table~\ref{tab:planetparameters}. We use Equations (4) and (6) from \cite{furlan2017} to account for dilution in the transit for the primary and secondary scenarios, respectively. When estimating the planet radius (R\textsubscript{p}) for our \ktwo derived parameters, we use the $\Delta Kep$ mag estimated in \S~\ref{subsec:hri}.  When estimating the planet radius from the \Spitzer fit, we estimated the $\Delta IRAC2$ band magnitude using our resolved component properties and the compiled $K_s-W2$ colors from \cite{pecaut2013}. We estimate $\Delta IRAC2$ = 0.89 $\pm$ 0.03 mag.  In each case, our planet radius and equilibrium temperature estimates assume Gaussian distributed uncertainties for the parameters from Table~\ref{tab:stellarparameters} and Table~\ref{tab:transit}.

We find that the \ktwo and \Spitzer planet radii estimated assuming the candidate orbits the secondary are more consistent than when assuming it orbits the primary. This result, along with the consistency between the stellar density estimated from the transit fit ($\rho_*$ = $25.70^{+6.77}_{-9.39}$~g~cm$^{-3}$); Table~\ref{tab:transit}) and the estimated density of the M3V companion based on resolved measurements ($\rho_*$ = 14.2 $\pm$ 5.0 ~g~cm$^{-3}$ ; Table~\ref{tab:stellarparameters}), and the significantly deeper transit in the the Spitzer IRAC2 band, provide evidence that the candidate orbits the secondary component in the system. 

Assuming the candidate transits the secondary, we applied the \vespa statistical planet validation tool to the system \citep{morton2015, montet2015}. We use the stellar parameters of the secondary provided in Table~\ref{tab:stellarparameters} as input. We also include the contrast curves from our resolved NIR imaging as additional input. \vespa returns a false positive probability (FPP) of 7.7 $\times$ 10\textsuperscript{-9} when using the folded \ktwo transit. This FPP indicates that the transiting signal is not a bound or background eclipsing binary and we consider the transiting body a validated planet.

We conclude, using evidence provided in this section, that the observed transit is caused by a planet on a 31.39 day orbital period and most likely occurs around the secondary M3V star. We calculate the weighted mean of the \ktwo and \Spitzer transit derived planet radii to arrive at $R_p$ = $1.9 \pm 0.3~R_{\oplus}$. Given both stellar and planet parameters provided in Table~\ref{tab:stellarparameters} and Table~\ref{tab:transit} respectively, we estimate the equilibrium temperature of the planet to be roughly 226K. We adopt the following nomenclature for this system: \primary is the primary M2V star, \secondary is the secondary M3V, and \planet is the planet.

\section{Conclusions} \label{sec:conclusions}
We present the discovery of a small, temperate (1.9 R$_{\Earth}$; 226 K) planet on a 31.39 day orbit likely around the lower-mass secondary of an M-dwarf binary system. The secondary is separated from the primary by a projected distance of $\approx$55 AU. This planetary system, \target, represents the third system identified by the citizen scientists of Exoplanet Explorers. 

\planet is an interesting target for several reasons beyond its discovery by citizen scientists. It resides in a moderate separation low-mass binary system and likely transits the secondary. Regardless of which star in the system it orbits, its equilibrium temperature places it in or near the habitable zone and its estimated radius places it in the ``Fulton gap" \citep{fulton2017, fulton2018, teske2018}, a likely transition zone between rocky super-Earths and volatile dominated sub-Neptunes. Thus, \planet has a radius that places it with other small planets that occur less frequently and it may still be undergoing atmospheric evolution.  \planet is similar to other known planetary systems where the planet orbits one component of a multiple system, for example Kepler-296AB \citep{barclay2015} and Kepler-444ABC \citep{dupuy2016}. However, this system hosts only a single detected transiting planet. Analyses of binary systems hosting transiting planets reveal that companions may have significant impacts on planet formation and evolution \citep{ziegler2018b, baszo2017}. 

Future resolved observations of the stars could place constraints on the semi-major axis, inclination, and eccentricity of their orbit to provide further insight on the effect of the companion on system formation and evolution \citep[e.g.][]{dupuy2016}. This is an interesting prospect given most known M dwarf systems are compact with small planets \citep[]{gillon2017, muirhead2015}, and the \target system hosts only a single planet with a relatively long  (31.39 day) period. \planet is also similar to other \ktwo discovered small, temperate planets transiting M dwarfs, such as K2-3d, K2-18b, and K2-9b \citep{crossfield2015, montet2015, benneke2017, schlieder2016} and is similar in size but significantly cooler than the well-known GJ1214b \citep{charbonneau2009}. Transit spectroscopy of \planet with future observatories could provide insight into atmosphere evolution of similar planets of significantly different equilibrium temperatures orbiting different host stars.

With the start of science operations of the Transiting Exoplanet Survey Satellite (TESS) mission \citep{ricker2015}, the stream of high precision photometric time series data will continue and increase in size. The role of citizen scientists will likely become even more crucial to the detection of interesting transiting exoplanets. Through continued engagement with the public via outreach and social media, we aim to foster continued interest in exoplanet citizen science and continue to validate interesting planetary systems which may otherwise be missed by automated software searches.
\\

\acknowledgements
We would like to acknowledge all other citizen scientists who were directly involved in flagging this system as well as those who continue to do so. This work, and hopefully many more in the future, was made possible by the Exoplanet Explorers project hosted on Zooniverse.org. Based on the responses from those citizen scientists who are credited as authors here, we encourage the practice of science teams reaching out to citizen scientists for all future discovery papers. 

We would additionally like to thank our anonymous referee for taking the time to review our report in great detail, which created a more complete picture of the system presented here.

MB acknowledges support from the North Carolina Space Grant Consortium. LA acknowledges support from NASA’s Minority University Research and Education Program Institutional Research Opportunity to the University of the Virgin Islands. EJG acknowledges support from the NSF graduate research fellowship program. MRK acknowledges support from the NSF Graduate Research Fellowship, grant No. DGE  1339067. BT acknowledges support from the National Science Foundation Graduate Research Fellowship under grant number DGE1322106 and NASA’s Minority University Research and Education Program. This work made use of the SIMBAD database (operated at CDS, Strasbourg, France); NASA's Astrophysics Data System Bibliographic Services; NASA's Exoplanet Archive and Infrared Science Archive; data products from theTwo Micron All Sky Survey (2MASS); the APASS database; the Digitized Sky Survey; and the Wide-Field Infrared Survey Explorer (WISE). This work has made use of data from the European Space Agency (ESA) mission {\it Gaia} (\url{https://www.cosmos.esa.int/gaia}), processed by the {\it Gaia} Data Processing and Analysis Consortium (DPAC, \url{https://www.cosmos.esa.int/web/gaia/dpac/consortium}). Funding for the DPAC has been provided by national institutions, in particular the institutions participating in the {\it Gaia} Multilateral Agreement. This paper includes data collected by the Kepler mission. Funding for the Kepler mission is provided by the NASA Science Mission directorate. Some of the data presented in this paper were obtained from the Mikulski Archive for Space Telescopes (MAST). STScI is operated by the Association of Universities for Research in Astronomy, Inc., under NASA contract NAS5-26555. Support for MAST for non-HST data is provided by the NASA Office of Space Science via grant NNX09AF08G and by other grants and contracts. Some of the data presented herein were obtained at the W. M. Keck Observatory, which is operated as a scientific partnership among the California Institute of Technology, the University of California and the National Aeronautics and Space Administration. The Observatory was made possible by the generous financial support of the W. M. Keck Foundation. The authors wish to recognize and acknowledge the very significant cultural role and reverence that the summit of Maunakea has always had within the indigenous Hawaiian community.  We are most fortunate to have the opportunity to conduct observations from this mountain.

\facilities{IRTF:3.0m (SpeX), Keck:I (HIRES), Keck:II (NIRC2), {\it Kepler}, {\it Spitzer}}

\vspace{5mm}

\software{k2phot \citep{petigura2015},~TERRA \citep{petigura2013a,petigura2013b},~EVEREST \citep{luger2017},~emcee \citep{emcee},~batman \citep{kreidberg2015},~vespa \citep{morton2015}}

\vspace{5mm}



\newpage
\begin{deluxetable}{l r r }[!ht]
\hspace{-1in}\tabletypesize{\scriptsize}
\tablecaption{  Stellar Parameters \label{tab:stellarparameters}}
\tablewidth{0pt}
\tablehead{
\colhead{Parameter} & \colhead{Value} & \colhead{Notes}
}
\startdata
\multicolumn{3}{c}{\em Identifying Information} \\
K2 ID & \target & \\
EPIC ID & 210693462 & \\ 
$\alpha$ R.A. (hh:mm:ss) & 03:41:46.43 & EPIC\\
$\delta$ Dec. (dd:mm:ss) & +18:16:08.0 & EPIC\\
$\mu_{\alpha}$ (mas~yr$^{-1}$) & $+186.1\pm 1.3$ & UCAC5\\
$\mu_{\delta}$ (mas~yr$^{-1}$) & $-61.2\pm1.2$ & UCAC5\\
Barycentric RV (km~s$^{-1}$)  &  $71.6 \pm 0.2$  & HIRES; This Work$^{a}$\\
Distance (pc) & 69.3 $\pm$ 0.4  & Gaia DR2$^{b}$\\
Age (Myr) & $\gtrsim$ 1 Gyr &  This Work\\
\\
\multicolumn{3}{c}{\em Blended Photometric Properties} \\
$B$ (mag) ..........  & $15.403 \pm 0.060$ & APASS DR9 \\
$V$ (mag) ..........  & $13.971 \pm 0.063$ & APASS DR9 \\
$g^{\prime}$ (mag) ..........  & $14.656 \pm 0.016$ & APASS DR9 \\
$r^{\prime}$ (mag) ..........  & $13.342 \pm 0.071$ & APASS DR9 \\
$Kep$ (mag) ..........        & $13.105$           & EPIC \\
$i^{\prime}$ (mag) ..........  & $12.456\pm 0.136$ & APASS DR9 \\
$J$ (mag) ..........  & $10.545 \pm 0.020$ & 2MASS \\
$H$ (mag) ..........  & $9.946 \pm 0.023$ & 2MASS \\
$K_s$ (mag) .........   & $9.724 \pm 0.018$ & 2MASS \\
$W1$ (mag) .........   & $9.595 \pm 0.024$ & ALLWISE \\
$W2$ (mag) .........   & $9.476 \pm 0.021$ & ALLWISE \\
$W3$ (mag) .........   & $9.394 \pm 0.038$ & ALLWISE \\
$W4$ (mag) .........   & $>8.750$ & ALLWISE \\
\\
\multicolumn{3}{c}{\em Individual Component Properties$^c$} \\
\underline{Primary}\\
Spectral Type .......... & M2V $\pm$ 1 & This Work \\
$B_p$ (mag) ..........     & $14.197 \pm 0.004$ & Gaia DR2 \\
$Kep$ (mag) ..........       & $13.46 \pm 0.09$ & This Work \\
$G$ (mag) ..........  & $13.309 \pm 0.001$ & Gaia DR2 \\
$R_p$ (mag) .......... & $11.982 \pm 0.003$ & Gaia DR2 \\
$J$ (mag) ..........  & $10.910 \pm 0.027$ & This Work \\
$H$ (mag) ..........  & $10.313 \pm 0.021$ & This Work \\
$K_s$ (mag) .........   & $10.092 \pm 0.023$ & This Work \\
$\lbrack $Fe/H$ \rbrack$ & $-0.29 \pm 0.09$ & HIRES; This Work$^{d}$\\
$M_*$ ($M_\odot$) .......... & $0.52 \pm 0.02$ & This Work\\
$R_*$ ($R_\odot$) .......... & $0.45 \pm 0.03$ & This Work\\
$T_{eff}$ (K) .......... & $3584 \pm 205$ & This Work\\
$log(L_*/L_\odot)$ .......... & $-1.49 \pm 0.02$ & This Work\\
log($g$) .......... & $4.85 \pm 0.03$ & This Work\\
$\rho$ (g cm$^{-3}$) .......... & $8.1 \pm 2.0$ & This Work\\
 & &  \\
\underline{Secondary}\\
Spectral Type .......... & M3V $\pm$ 1 & This Work \\
$Kep$ (mag) ..........       & $14.49\pm 0.10$ & This Work \\
$G$ (mag) ..........  & $14.545\pm 0.002$ & Gaia DR2 \\
$J$ (mag) ..........  & $11.907 \pm 0.027$ & This Work \\
$H$ (mag) ..........  & $11.303 \pm 0.021$ & This Work \\
$K_s$ (mag) .........   & $11.079 \pm 0.023$ & This Work \\
$\lbrack $Fe/H$ \rbrack$ & $-0.21 \pm 0.09$ & HIRES; This Work$^{d}$\\
$M_*$ ($M_\odot$) .......... & $0.33 \pm 0.02$ & This Work\\
$R_*$ ($R_\odot$) .......... & $0.32 \pm 0.03$ & This Work\\
$T_{eff}$ (K) .......... & $3341 \pm 276$ & This Work\\
$log(L_*/L_\odot)$ .......... & $-1.93 \pm 0.02$ & This Work\\
log($g$) .......... & $4.96 \pm 0.02$ & This Work\\
$\rho$ (g cm$^{-3}$) .......... & $14.2 \pm 5.0$ & This Work\\
\enddata

\tablenotetext{}{$^a$Weighted mean of the HIRES measured barycentric RVs of two blended spectra and a partially resolved spectrum of the secondary, see \S~\ref{subsec:hires},$^b$Weigthed mean of the Gaia DR2 distances of the primary and secondary; $^c$Stellar parameters from this section were used in the planet transit analyses; $^{d}$From HIRES spectroscopy using SpecMatch-Emp. The metallicity of the primary was measured from a blended spectrum containing light from the secondary; Gaia - \citep[]{gaia2018}, UCAC5 - \citep[]{zacharias2016}, APASS DR9 - \citep[]{henden2016}, EPIC - \citep[]{huber2016}, 2MASS - \citep[]{cutri2003}, ALLWISE - \citep[]{cutri2013}}
\end{deluxetable}

\begin{deluxetable}{lcr}[!ht]
\tabletypesize{\footnotesize}
\tablecaption{Transit parameters. \label{tab:transit}}
\tablehead{
\colhead{Parameter} & \colhead{Unit} & \colhead{Value}
}
\startdata
T\textsubscript{0}-2454833 &  BJD\textsubscript{TDB}      &   $2230.402366^{+0.001162}_{-0.001107} $ \\
P               & days      &   $31.393463^{+0.000067}_{-0.000069}$ \\
$a$             & $R_\star$ &   $110.2^{+8.9}_{-15.5}$ \\
b               & --- &         $0.37^{+0.23}_{-0.25}$ \\
i               & deg. &        $89.81^{+0.13}_{-0.17}$ \\
$R_{p,K}$       & $R_\star$ &   $0.0356^{+0.0012}_{-0.0010}$ \\
$R_{p,S}$       & $R_\star$ &   $0.0487^{+0.0030}_{-0.0031}$ \\
$u_{1,K}$       & --- &         $0.314^{+0.055}_{-0.057}$ \\
$u_{2,K}$       & --- &         $0.397^{+0.034}_{-0.033}$ \\
$u_{1,S}$       & --- &         $0.009^{+0.011}_{-0.007}$ \\
$u_{2,S}$       & --- &         $0.192^{+0.014}_{-0.014}$ \\
log($\sigma_K$) & --- &         $-8.88^{+0.13}_{-0.12}$ \\
log($\sigma_S$) & --- &         $-6.88^{+0.04}_{-0.04}$ \\
$\rho_{\star,\mathrm{circ}}$    & g\,cm$^{-3}$ &   $25.70^{+6.77}_{-9.39}$ \\
$T_{14}$        & days &         $0.088^{+0.003}_{-0.004}$ \\
$T_{23}$        & days &         $0.079^{+0.003}_{-0.003}$ \\
shape           & --- &         $0.91^{+0.01}_{-0.03}$ \\
$R_{p,\mathrm{max}}$     & $R_\star$ &   $0.049^{+0.019}_{-0.006}$ \\
\enddata
\tablecomments{The subscripts $K$ and $S$ refer to the \Kepler and \Spitzer 4.5\,\micron\, bandpasses, respectively. The parameter ``shape'' is the ratio of $T_{23}$ to $T_{14}$, where values close to unity indicate a ``box-shaped'' transit caused by a small occulting body. The parameter $R_{p,max}$ corresponds to the maximum planetary radius (in units of the stellar radius) allowed by the transit geometry. log($\sigma_K$) and log($\sigma_S$) represent the width of the zero-mean Gaussian errors. } 
\end{deluxetable}

\begin{deluxetable}{l r r }[!ht]
\tabletypesize{\footnotesize}
\tablecaption{Planet Parameters \label{tab:planetparameters}}
\tablehead{
\colhead{Parameter} & \colhead{Unit} & \colhead{Value} 
}
\startdata
\underline{Primary}\\
$R_{p,K}$  & $R_{\Earth}$  & 2.06  $\pm$ 0.16 \\
$R_{p,S}$  & $R_{\Earth}$  & 2.86  $\pm$ 0.27 \\
$a$        & AU            & 0.231 $\pm$ 0.03  \\
$T_{eq}$   & K             & 242.85$\pm$ 19.8  \\
 & &  \\
\underline{Secondary}\\
$R_{p,K}$  & $R_{\Earth}$  & 1.70   $\pm$ 0.36 \\
$R_{p,S}$  & $R_{\Earth}$  & 2.23   $\pm$ 0.47 \\
$a$        & AU            & 0.164  $\pm$ 0.03  \\
$T_{eq}$   & K             & 226.36 $\pm$ 22.3  \\
\enddata

\tablecomments{The subscripts $K$ and $S$ refer to the \Kepler and \Spitzer 4.5\,\micron\, bandpasses, respectively.}
\end{deluxetable}


\begin{thebibliography}{}

\bibitem[Abell(1966)]{abell1966} Abell, G.~O.\ 1966, \apj, 144, 259 

\bibitem[Aigrain et al.(2016)]{aigrain2016} Aigrain, S., Parviainen, H., \& Pope, B.~J.~S.\ 2016, \mnras, 459, 2408

\bibitem[Altmann et al.(2017)]{altmann2017} Altmann, M., Roeser, S., Demleitner, M., Bastian, U., \& Schilbach, E.\ 2017, \aap, 600, L4

\bibitem[Bailer-Jones et al.(2018)]{bailer2018} Bailer-Jones, C.~A.~L., Rybizki, J., Fouesneau, M., Mantelet, G., \& Andrae, R.\ 2018, \aj, 156, 58 

\bibitem[Barclay et al.(2015)]{barclay2015} Barclay, T., Quintana, E.~V., Adams, F.~C., et al.\ 2015, \apj, 809, 7 

\bibitem[Bazs{\'o} et al.(2017)]{baszo2017} Bazs{\'o}, {\'A}., Pilat-Lohinger, E., Eggl, S., et al.\ 2017, \mnras, 466, 1555 



\bibitem[{{Beichman} {et~al.}(2016){Beichman}, {Livingston}, {Werner},
  {Gorjian}, {Krick}, {Deck}, {Knutson}, {Wong}, {Petigura}, {Christiansen},
  {Ciardi}, {Greene}, {Schlieder}, {Line}, {Crossfield}, {Howard}, \&
  {Sinukoff}}]{2016ApJ...822...39B}
{Beichman}, C., {Livingston}, J., {Werner}, M., {et~al.} 2016, \apj, 822, 39

\bibitem[Benedict et al.(2016)]{benedict2016} Benedict, G.~F., Henry, T.~J., Franz, O.~G., et al.\ 2016, \aj, 152, 141 

\bibitem[Benneke et al.(2017)]{benneke2017} Benneke, B., Werner, M., Petigura, E., et al.\ 2017 \apj, 834, 187

\bibitem[Boyajian et al.(2016)]{boyajian2016} Boyajian, T.~S., LaCourse, D.~M., Rappaport, S.~A., et al.\ 2016, \mnras, 457, 3988 

\bibitem[Burgasser et al.(2002)]{burgasser2002} Burgasser, A.~J., Kirkpatrick, J.~D., Brown, M.~E., et al.\ 2002, \apj, 564, 421 

\bibitem[Carpenter(2001)]{carpenter2001} Carpenter, J.~M.\ 2001, \aj, 121, 2851 

\bibitem[Charbonneau et al. (2009)]{charbonneau2009} Charbonneau, D., Berta, Z. K., Irwin, J., et al. \ 2009, Nature, 462, 891

\bibitem[Christiansen et al.(2018)]{christiansen2018} Christiansen, J. ~L., Crossfield, I. ~J. ~M., Barenstein, G., et al. \ 2018, \aj, 155, 2

\bibitem[Ciardi et al.(2015)]{ciardi2015} Ciardi, D.~R., Beichman, C.~A., Horch, E.~P., \& Howell, S.~B.\ 2015, \apj, 805, 16

\bibitem[Ciardi et al. (2018)]{ciardi2018} Ciardi, D., Crossfield, I.~J.~M., Feinstein, A.~D., et al.\ 2018, \apj, 155, 10

\bibitem[Claret et al. (2012)]{claret2012} Claret, A., Hauschildt, P.~H., Witte, S.\ 2012, A\&A, 546, A14

\bibitem[{{Claret} {et~al.}(2012){Claret}, {Hauschildt}, \&
  {Witte}}]{2012yCat..35460014C}
{Claret}, A., {Hauschildt}, P.~H., \& {Witte}, S. 2012, VizieR Online Data
  Catalog, 354

\bibitem[Crossfield et al.(2015)]{crossfield2015} Crossfield, I.~J.~M., Petigura, E., Schlieder, J.~E., et al.\ 2015, \apj, 804, 10 

\bibitem[Crossfield et al.(2016)]{crossfield2016} Crossfield, I.~J.~M., Ciardi, D.~R., Petigura, E.~A., et al.\ 2016, \apjs, 226, 7

\bibitem[Crossfield et al.(2018)]{crossfield2018} Crossfield, I.~J.~M., Guerrero, N., David, T., et al.\ 2018, arXiv:1806.03127 


\bibitem[Cushing et al.(2004)]{cushing2004} Cushing, M. ~C., Vacca, W. ~D., \& Rayner, J. ~T. \ 2004, \pasp, 116, 362

\bibitem[Cutri et al. (2003)]{cutri2003} Cutri, R. ~M.,  Skrutskie, M. ~F., van Dyk, S. et al.\ 2003, 2MASS All Sky Catalog of point sources, 2246, 0

\bibitem[Cutri et al. (2014)]{cutri2013} Cutri, R. ~M., et al. \ 2014, VizieR Online Data Catalog, 2328, 0

\bibitem[David et al.(2018)]{david2018} David, T.~J., Crossfield, I.~J.~M., Benneke, B., et al.\ 2018, arXiv:1803.05056 

\bibitem[{{Deming} {et~al.}(2015){Deming}, {Knutson}, {Kammer}, {Fulton},
  {Ingalls}, {Carey}, {Burrows}, {Fortney}, {Todorov}, {Agol}, {Cowan},
  {Desert}, {Fraine}, {Langton}, {Morley}, \& {Showman}}]{2015ApJ...805..132D}
{Deming}, D., {Knutson}, H., {Kammer}, J., {et~al.} 2015, \apj, 805, 132

\bibitem[Dupuy et al.(2016)]{dupuy2016} Dupuy, T.~J., Kratter, K.~M., Kraus, A.~L., et al.\ 2016, \apj, 817, 80 

\bibitem[Dressing et al.(2017)]{dressing2017a} Dressing, C.~D., Newton, E.~R., Schlieder, J.~E., et al.\ 2017, \apj, 836, 167

\bibitem[Dressing et al.(2018)]{dressing2018sub} Dressing, C.~D., Hardegree-Ullman, K., Schlieder, J.~E., et al.\ 2018, \apj, submitted

\bibitem[Evans et al.(2018)]{Evans2018} Evans, D.~W., Riello, M., De Angeli, F., et al.\ 2018, arXiv:1804.09368 

\bibitem[{{Fazio} {et~al.}(2004){Fazio}, {Hora}, {Allen}, {Ashby}, {Barmby},
  {Deutsch}, {Huang}, {Kleiner}, {Marengo}, {Megeath}, {Melnick}, {Pahre},
  {Patten}, {Polizotti}, {Smith}, {Taylor}, {Wang}, {Willner}, {Hoffmann},
  {Pipher}, {Forrest}, {McMurty}, {McCreight}, {McKelvey}, {McMurray}, {Koch},
  {Moseley}, {Arendt}, {Mentzell}, {Marx}, {Losch}, {Mayman}, {Eichhorn},
  {Krebs}, {Jhabvala}, {Gezari}, {Fixsen}, {Flores}, {Shakoorzadeh}, {Jungo},
  {Hakun}, {Workman}, {Karpati}, {Kichak}, {Whitley}, {Mann}, {Tollestrup},
  {Eisenhardt}, {Stern}, {Gorjian}, {Bhattacharya}, {Carey}, {Nelson},
  {Glaccum}, {Lacy}, {Lowrance}, {Laine}, {Reach}, {Stauffer}, {Surace},
  {Wilson}, {Wright}, {Hoffman}, {Domingo}, \& {Cohen}}]{2004ApJS..154...10F}
{Fazio}, G.~G., {Hora}, J.~L., {Allen}, L.~E., {et~al.} 2004, \apjs, 154, 10

\bibitem[Fischer et al.(2012)]{fischer2012} Fischer, D. ~A., Schwamb, M. ~E., Schawinski, K., et al. \ 2012, \mnras, 419, 2900

\bibitem[{{Foreman-Mackey} {et~al.}(2013){Foreman-Mackey}, {Hogg}, {Lang}, \&
  {Goodman}}]{emcee}
{Foreman-Mackey}, D., {Hogg}, D.~W., {Lang}, D., \& {Goodman}, J. 2013, PASP,
  125, 306

\bibitem[Fulton et al. (2017)]{fulton2017} Fulton, B.~J., Petigura, E.~A., Howard, A.~W., et al. \ 2017 \apj 154, 3

\bibitem[Fulton \& Petigura(2018)]{fulton2018} Fulton, B.~J., \& Petigura, E.~A.\ 2018, arXiv:1805.01453 


\bibitem[Furlan et al.(2017)]{furlan2017} Furlan, E., Ciardi, D.~R., Everett, M.~E., et al.\ 2017, \aj, 153, 71

\bibitem[Gaia Collaboration et al.(2018)]{gaia2018} Gaia Collaboration, Brown, A.~G.~A., Vallenari, A., et al.\ 2018, arXiv:1804.09365 

\bibitem[Gies et al.(2013)]{gies2013} Gies, D.~R., Guo, Z., Howell, S.~B., et al.\ 2013, \apj, 775, 64 

\bibitem[Gillon et al. (2017)]{gillon2017} Gillon, M., Triaud, A.~H.~M.~J.,Demory, B.~O. et al. \ 2017, Nature, 542, 7642

\bibitem[Hawley et al.(2002)]{hawley2002} Hawley, S.~L., Covey, K.~R., Knapp, G.~R., et al.\ 2002, \aj, 123, 3409 

\bibitem[Henden et al. (2016)]{henden2016} Henden, A. A., Templeton, M., Terrell, D., et al. 2016, VizieR
Online Data Catalog, 2336

\bibitem[Howard et al.(2010)]{howard2010} Howard, A.~W., Johnson, J.~A., Marcy, G.~W., et al.\ 2010, \apj, 721, 1467

\bibitem[Howell et al.(2012)]{howell2012} Howell, S.~B., Rowe, J.~F., Bryson, S.~T., et al.\ 2012, \apj, 746, 123 

\bibitem[Howell et al.(2014)]{howell2014} Howell, S. ~B., Sobeck, C., Haas, M., et al. \ 2014, \pasp, 126, 398

\bibitem[Huber et al.(2016)]{huber2016} Huber, D., Bryson, S.~T., Haas, M.~R., et al.\ 2016, \apjs, 224, 2 

\bibitem[Kirkpatrick et al.(2000)]{kirkpatrick2000} Kirkpatrick, J.~D., Reid, I.~N., Liebert, J., et al.\ 2000, \aj, 120, 447 

\bibitem[{{Kipping}(2013)}]{2013MNRAS.435.2152K}
{Kipping}, D.~M. 2013, \mnras, 435, 2152

\bibitem[{{Knutson} {et~al.}(2012){Knutson}, {Lewis}, {Fortney}, {Burrows},
  {Showman}, {Cowan}, {Agol}, {Aigrain}, {Charbonneau}, {Deming}, {D{\'e}sert},
  {Henry}, {Langton}, \& {Laughlin}}]{2012ApJ...754...22K}
{Knutson}, H.~A., {Lewis}, N., {Fortney}, J.~J., {et~al.} 2012, \apj, 754, 22

\bibitem[Kolbl et al.(2015)]{kolbl2015} Kolbl, R., Marcy, G. ~W., Isaacson, H., \& Howard, A. ~W. \ 2015, \aj, 149, 18

\bibitem[Kreidberg (2015)]{kreidberg2015} Kreidberg, L.\ 2015, \pasp, 127, 1161

\bibitem[{{Kreidberg}(2015)}]{2015PASP..127.1161K}
{Kreidberg}, L. 2015, \pasp, 127, 1161

\bibitem[L{\'e}pine et al.(2003)]{lepine2003} L{\'e}pine, S., Rich, R. ~M., \& Shara, M. ~M. \ 2003, \aj, 125, 3

\bibitem[L{\'e}pine et al.(2013)]{lepine2013} L{\'e}pine, S., Hilton, E.~J., Mann, A.~W., et al.\ 2013, \aj, 145, 102

\bibitem[Lindegren et al.(2018)]{Lindegren2018} Lindegren, L., Hernandez, J., Bombrun, A., et al.\ 2018, arXiv:1804.09366 

\bibitem[Lintott et al.(2008)]{lintott2008} Lintott, C. ~J., Schawinski, K., Slosar, A., et al. \ 2008, \mnras, 389, 1179

\bibitem[Luger et al.(2016)]{luger2016} Luger, R., Agol, E., Kruse, E., et al.\ 2016, \aj, 152, 100

\bibitem[Luger et al.(2017)]{luger2017} Luger, R., Kruse, E., Foreman-Mackey, D., Agol, E., \& Saunders, N.\ 2017, arXiv:1702.05488

\bibitem[Mandel \& Agol (2002)]{mandel2002} Mandel, K., \& Agol, E. 2002, \apj, 580, L171

\bibitem[Mann et al.(2013a)]{mann2013a} Mann, A.~W., Brewer, J.~M., Gaidos, E., L{\'e}pine, S., \& Hilton, E.~J.\ 2013, \aj, 145, 52

\bibitem[Mann et al.(2013b)]{mann2013b} Mann, A.~W., Gaidos, E., Ansdell, M \ 2013, \aj, 779, 188

\bibitem[Mann et al.(2015)]{mann2015} Mann, A.~W., Feiden, G.~A., Gaidos, E., Boyajian, T., \& von Braun, K.\ 2015, \apj, 804, 64

\bibitem[Mann et al.(2016)]{mann2016} Mann, A.~W., Feiden, G.~A., Gaidos, E., Boyajian, T., \& von Braun, K.\ 2016, \apj, 819, 87

\bibitem[Marcy et al.(2008)]{marcy2008} Marcy, G.~W., Butler, R.~P., Vogt, S.~S., et al.\ 2008, Physica Scripta Volume T, 130, 014001

\bibitem[Mayo et al.(2018)]{mayo2018} Mayo, A.~W., Vanderburg, A., Latham, D.~W., et al.\ 2018, \aj, 155, 136

\bibitem[Montet et al.(2015)]{montet2015} Montet, B.~T., Morton, T.~D., Foreman-Mackey, D., \ 2015, \apj, 809, 25

\bibitem[Morton(2015)]{morton2015} Morton, T.~D.\ 2015, Astrophysics Source Code Library, ascl:1503.011 

\bibitem[Muirhead et al.(2015)]{muirhead2015} Muirhead, P.~S., Mann, A.~W., Vanderburg, A., et al. \ 2015, \apj, 801, 18

\bibitem[NASA Exoplanet Archive(2018)]{NEA} NASA Exoplanet Archive, 2018, Update 2018 February 1

\bibitem[Newton et al.(2015)]{newton2015} Newton, E.~R., Charbonneau, D., Irwin, J., \& Mann, A.~W.\ 2015, \apj, 800, 85

\bibitem[{Newville {et~al.}(2014)Newville, Stensitzki, Allen, \&
  Ingargiola}]{newville_2014_11813}
Newville, M., Stensitzki, T., Allen, D.~B., \& Ingargiola, A. 2014, {LMFIT:
  Non-Linear Least-Square Minimization and Curve-Fitting for Python¶}, , ,
  doi:10.5281/zenodo.11813.
\newblock \url{https://doi.org/10.5281/zenodo.11813}

\bibitem[Pecaut \& Mamajek(2013)]{pecaut2013} Pecaut, M.~J., \& Mamajek, E.~E.\ 2013, \apjs, 208, 9 

\bibitem[Petigura et al.(2013a)]{petigura2013a} Petigura, E.~A., Marcy, G.~W., \& Howard, A.~W.\ 2013, \apj, 770, 69

\bibitem[Petigura et al.(2013b)]{petigura2013b} Petigura, E.~A., Howard, A.~W., \& Marcy, G.~W.\ 2013, Proceedings of the National Academy of Science, 110, 19273

\bibitem[Petigura et al.(2015)]{petigura2015} Petigura, E.~A., Schlieder, J.~E., Crossfield, I.~J.~M., et al.\ 2015, \apj, 811, 102

\bibitem[Petigura et al.(2018)]{petigura2018} Petigura, E.~A., Crossfield, I.~J.~M., Isaacson, H., et al.\ 2018, \aj, 155, 21

\bibitem[{{Pont} {et~al.}(2006){Pont}, {Zucker}, \&
  {Queloz}}]{2006MNRAS.373..231P}
{Pont}, F., {Zucker}, S., \& {Queloz}, D. 2006, \mnras, 373, 231

\bibitem[Rayner et al.(2003)]{rayner2003} Rayner, J. ~T., Toomey, D. ~W., Onaka, P. ~M., et al. \ 2003, \pasp, 115, 362

\bibitem[Rayner et al.(2004)]{rayner2004} Rayner, J.~T., Onaka, P.~M., Cushing, M.~C., \& Vacca, W.~D.\ 2004, \procspie, 5492, 1498

\bibitem[Rayner et al.(2009)]{rayner2009} Rayner, J. ~T., Cushing, M. ~C., \& Vacca, W. ~D. \ 2009, \apj, 185, 289

\bibitem[Ricker et al.(2015)]{ricker2015} Ricker, G.~R., Winn, J.~N., Vanderspek, R., et al.\ 2015, Journal of Astronomical Telescopes, Instruments, and Systems, 1, 014003 

\bibitem[Riello et al.(2018)]{Riello2018} Riello, M., De Angeli, F., Evans, D.~W., et al.\ 2018, arXiv:1804.09367 

\bibitem[Rizzuto et al.(2018)]{rizzuto2018} Rizzuto, A.~C., Vanderburg, A., Mann, A.~W., et al.\ 2018, arXiv:1808.07068 

\bibitem[Rojas-Ayala et al.(2012)]{rojas2012} Rojas-Ayala, B., Covey, K.~R., Muirhead, P.~S., \& Lloyd, J.~P.\ 2012, \apj, 748, 93

\bibitem[Schlafly \& Finkbeiner(2011)]{sf2011} Schlafly, E.~F., \& Finkbeiner, D.~P.\ 2011, \apj, 737, 103 

\bibitem[Schlieder et al., (2016)]{schlieder2016} Schlieder, J.~E., Crossfield, I.~J.~M., Petigura, E.~A., et al\ 2016, \apj, 818, 87

\bibitem[Schmitt et al.(2014)]{schmitt2014} Schmitt, J.~R., Agol, E., Deck, K.~M., et al.\ 2014, \apj, 795, 167 

\bibitem[Schwamb et al.(2012)]{schwamb2012} Schwamb, M.~E., Lintott, C.~J., Fischer, D.~A., et al.\ 2012, \apj, 754, 129 

\bibitem[Teske et al.(2018)]{teske2018} Teske, J.~K., Ciardi, D.~R., Howell, S.~B., Hirsch, L.~A., \& Johnson, R.~A.\ 2018, arXiv:1804.10170

\bibitem[Vacca et al.(2003)]{vacca2003} Vacca, W. ~D., Cushing, M. ~C., \& Rayner, J. ~T. \ 2003, \pasp, 115, 389

\bibitem[Vanderburg \& Johnson(2014)]{vanderburg2014} Vanderburg, A., \& Johnson, J.~A.\ 2014, \pasp, 126, 948 

\bibitem[Vogt et al.(1994)]{vogt1994} Vogt, S. ~S., Allen, S. ~L., Bigelow, B. ~C., et al. \ 1994, in Proc. SPIE, Vol. 2198, Instrumentation in Astronomy VIII, ed. D. L. Crawford \& E. R. Craine, 362

\bibitem[Wang et al.(2013)]{wang2013} Wang, J., Fischer, D.~A., Barclay, T., et al.\ 2013, \apj, 776, 10 

\bibitem[{{Winn} {et~al.}(2008){Winn}, {Holman}, {Torres}, {McCullough},
  {Johns-Krull}, {Latham}, {Shporer}, {Mazeh}, {Garcia-Melendo}, {Foote},
  {Esquerdo}, \& {Everett}}]{2008ApJ...683.1076W}
{Winn}, J.~N., {Holman}, M.~J., {Torres}, G., {et~al.} 2008, \apj, 683, 1076

\bibitem[Yee et al.(2017)]{yee2017} Yee, S. ~W., Petigura, E. ~A., \& von Braun, K. \ 2017, \apj, 836, 77

\bibitem[Yu et al.(2018)]{yu2018} Yu, L., Crossfield, I.~J.~M., Schlieder, J.~E., et al.\ 2018, arXiv:1803.04091 

\bibitem[Zacharias et al. (2017)]{zacharias2016} Zacharias, N., Finch, C., Frouard, J., et al. \ 2017, \aj, 153, 4 

\bibitem[Ziegler et al.(2018)]{ziegler2018a} Ziegler, C., Law, N.~M., Baranec, C., et al.\ 2018, arXiv:1806.10142 

\bibitem[Ziegler et al.(2018)]{ziegler2018b} Ziegler, C., Law, N.~M., Baranec, C., et al.\ 2018, \aj, 156, 83 



\end{thebibliography}
\end{document}